%% file: revise2_dynmri_lrcs.tex
\documentclass[journal, 10pt]{IEEEtran} 

\usepackage{amsmath,graphicx,bm,bbm,algorithm,color} 
\usepackage{amsmath, mathtools, amsfonts, bm, amssymb, amsthm,graphicx, caption, subcaption,multirow}
\usepackage{algorithm} 
\usepackage[colorlinks=true, allcolors=blue]{hyperref}

\usepackage{algpseudocode,matlab-prettifier}

\newcommand{\ik}{{ki}}

\renewcommand{\a}{\bm{a}}

\newcommand{\x}{\bm{x}}

\newcommand{\e}{\bm{e}}

\renewcommand{\b}{\bm{b}}
\newcommand{\y}{\bm{y}}
\newcommand{\w}{\bm{w}}

\newcommand{\U}{{\bm{U}}}

\newcommand{\X}{\bm{X}}
\newcommand{\R}{{\bm{R}}}
\newcommand{\B}{\bm{B}}

\newcommand{\I}{\bm{I}}
\newcommand{\Y}{\bm{Y}}

\newcommand{\s}{\bm{s}}

\newcommand{\xhat}{\bm{{x}}}
\newcommand{\bhat}{\bm{{b}}}
\newcommand{\Uhat}{\U}
\newcommand{\Bhat}{\B}
\newcommand{\Xhat}{\X}

\newcommand{\Z}{{\bm{Z}}}

\newcommand{\D}{{\bm{D}}}

\newcommand{\F}{{\bm{F}}}

\newcommand{\z}{\bm{z}}
\newcommand{\indic}{\mathbbm{1}}

\newcommand{\SE}{\mathrm{SD}}
\newcommand{\SEF}{\SE}
\newcommand{\dist}{\mathrm{dist}}

\newcommand{\A}{\bm{A}}
\renewcommand{\H}{\bm{H}}

\newcommand{\M}{\bm{M}}

\renewcommand{\S}{\bm{S}}

\newcommand{\Xstar}{{\X^*}}
\newcommand{\xstar}{\x^*}
\newcommand{\qreq}{\overset{\mathrm{QR}}=} 

\newtheorem{theorem}{Theorem}[section]

\newcommand{\ds}{\displaystyle}
\renewcommand{\forall}{\text{ for all }}

\newcommand{\Subsection}[1]{ \vspace{-0.1in} \subsection{#1}  \vspace{-0.02in} }   
\renewcommand{\subsubsection}[1]{\noindent {\em{#1. }}}

\newcommand{\vsm}{\vspace{-0.1in}}

\newcommand{\estar}{\e^*}
\newcommand{\zstar}{\z^*}
\newcommand{\Zstar}{\Z^*}
\newcommand{\ty}{\tilde{\y}}

\newcommand{\zbar}{\bar{\z}}
\newcommand{\zstarbar}{\bar{\zstar}}

\newcommand{\sF}{\mathcal{F}_{row}}
\newcommand{\sA}{\mathcal{A}}
\newcommand{\Emat}{\bm{E}}
\newcommand{\Mmat}{\bm{M}}
\newcommand{\Sstar}{\S^*}
\newcommand{\Estar}{\Emat^*}

\newcommand{\rhat}{{\hat{r}}}

\newcommand{\mbar}{\bar{m}}

\newcommand{\bea}{\begin{eqnarray}} \newcommand{\eea}{\end{eqnarray}}
\newcommand{\ben}{\begin{enumerate}} \newcommand{\een}{\end{enumerate}}

\newcommand{\cred}{} \newcommand{\cblue}{} 
\title{Fast Low Rank column-wise Compressive Sensing \\ for Accelerated Dynamic MRI}
\author{Silpa Babu*, Sajan Goud Lingala**, Namrata Vaswani* \\  
*ECE dept, Iowa State University, USA \\ **BME dept, University of Iowa, USA
\thanks{This work was partially supported by NSF grant CIF-2115200. An early version of this work with the same title was presented at ICASSP 2022 \cite{lrpr_gdmin_mri}.}}

\begin{document}
%
\maketitle

\renewcommand{\bhat}{\b}
\renewcommand{\xhat}{\x}
\renewcommand{\Xhat}{\X}


\begin{abstract}
This work develops a novel set of algorithms, alternating Gradient Descent (GD) and minimization for MRI (altGDmin-MRI1 and altGDmin-MRI2),  for accelerated dynamic MRI by assuming an approximate low-rank (LR) model on the matrix formed by the vectorized images of the sequence. The LR model itself is well-known in the MRI literature; our contribution is the novel GD-based algorithms which are much faster, memory-efficient, and `general' compared with existing work; and careful use of a 3-level hierarchical LR model.  By `general', we mean that, with a single choice of parameters, our method provides accurate reconstructions for multiple accelerated dynamic MRI applications, multiple sampling rates and sampling schemes. We show that our methods outperform many of the popular existing approaches while also being faster than all of them, on average. This claim is based on comparisons on 8 different retrospectively undersampled multi-coil dynamic MRI applications, sampled using either 1D Cartesian or 2D pseudo-radial undersampling, at multiple sampling rates. Evaluations on some prospectively undersampled datasets are also provided.
Our second contribution is a mini-batch subspace tracking extension that can process new measurements and return reconstructions within a short delay after they arrive. The recovery algorithm itself is also faster than its batch counterpart.
\end{abstract}

{\em Index terms: } low-rank, compressed sensing, MRI

\section{Introduction}
Dynamic Magnetic Resonance Imaging (MRI) is a powerful imaging modality to non-invasively capture time evolving phenomena in the human body, such as the beating heart, motion of vocal tract during speaking,  or dynamics of contrast uptake in brain. 
A long standing challenge in MRI is its slow imaging speed which restricts its full potential in the achievable spatial or temporal resolution. 
From a signal processing standpoint, in MRI, one measures the 2D discrete Fourier transform (FT) of a slice of the organ being imaged, one FT coefficient (or one line of coefficients) at a time. This makes the imaging slow. 
Accelerated/undersampled/compressive MRI is one of the key practical applications where Compressive Sensing (CS) ideas have been extensively used for speeding up the scan.
This includes both work that uses traditional (sparse) CS \cite{sparsemri,sparsedynamicMRI} for single image MRI, as well as later work that relies on the low-rank (LR) assumption, e.g., \cite{liang2007spatiotemporal,lingala2011accelerated,kt_faster_LR_fmri,kt_faster_LR_fmri_2}. 

\subsection{Our Contributions}\label{contrib}
This work develops a fast, memory-efficient, and {\em `general'} algorithm, called altGDmin-MRI, for accelerated dynamic MRI by assuming an approximate LR model on the matrix formed by the vectorized images of the sequence. In analogy with traditional (sparse) CS, we refer to the problem of reconstruction with this modeling assumption as approximate LR column-wise CS (LRcCS). We should mention here that LRcCS based models have been extensively used in past work in MRI \cite{liang2007spatiotemporal,pedersen2009k,lingala2011accelerated,zhao2012image,kt_faster_LR_fmri,kt_faster_LR_fmri_2}. Our contribution is a novel set of algorithms that assume a 3-level hierarchical LR model; and extensive experiments to demonstrate that these are both fast and  ``general'' (with a single set of parameters, these provide good enough reconstructions for many different MRI applications, sampling schemes and rates). 
Our methods are modifications of a fast GD-based algorithm that was developed and theoretically analyzed in our recent work~\cite{lrpr_gdmin}.

Our second contribution is mini-batch and online ``subspace tracking'' extensions of altGDmin-MRI that can process new measurements and return reconstructions after a much shorter data acquisition delay than the full batch solution. The online extension needs to be initialized with a mini-batch of measurements, but, after that, it returns the reconstruction as soon as a new frame of measurements arrives. The reconstruction algorithms are also faster and more memory-efficient than their batch counterpart, but with a gradual degradation in quality as batch size is reduced.

Reconstruction algorithm speed is an important concern in applications needing low latency such as real-time interactive MRI, interventional MRI, or biofeedback imaging. Moreover, immediate reconstructions can also allow for on the fly identification of certain artifacts, which can be immediately corrected (e.g., adjusting center frequency to minimize off-resonance artifacts, re-scanning if subject experiences sudden motion such as cough, etc). Finally, even in offline settings, if a reconstruction can be obtained without making the patient wait too long, it would considerably improve clinical workflow and overall throughput.  Fast reconstructions therefore help reduce the overhead cost associated with re-scheduling patient scans.%






\subsection{Existing work}

\subsubsection{Provable LRcCS solutions with only simulation experiments}
There are three existing provable solutions to the LRcCS problem. The first is an Alternating Minimization (AltMin) solution that is designed to solve a generalization of LRcCS \cite{lrpr_icml,lrpr_best}, and hence also solves LRcCS. The second studies a convex relaxation called mixed norm minimization (MixedNorm) \cite{lee2019neurips}. The third is the altGDmin solution \cite{lrpr_gdmin} that we modify in the current work.
The convex solution is extremely slow, both theoretically and experimentally; and it has a worse sample complexity in regimes of practical interest; see Table \ref{theory_algos1} and see \cite{lrpr_gdmin}.
The AltMin solution is faster than the convex one, but still significantly slower than AltGDmin \cite{lrpr_gdmin}.
All the proven theoretical guarantees are for random Gaussian measurements (each entry of each $\A_k$ is an independent identically distributed standard Gaussian) but, as in case of (sparse) CS \cite{cscompressible,donoho,sparsemri}, we expect the qualitative implications to remain true also for MRI which involves use of undersampled Fourier measurements.



\subsubsection{MRI literature: LR and sparsity based approaches}
Since the work on CS in the early 2000s there has been extensive work on exploiting sparsity of the image or of the sequence in different dictionaries and bases in order to enable accelerated MRI, e.g., see \cite{sparsemri,feng2016xd}  and follow-up work.
For settings where joint reconstruction of a set of similar images is needed, LR is a more flexible model since it does not require knowledge of the sparsifying basis or dictionary. MR images change slowly over time and hence are well-modeled as being approximately LR. Prior LR model based solutions from the MRI literature include \cite{gupta2001dynamic,pedersen2009k,lingala2011accelerated,zhao2012image,kt_faster_LR_fmri,kt_faster_LR_fmri_2} 
can be classified into two broad categories: (a)  methods which enforce the LR constraint explicitly, e.g., via explicit estimation of the temporal subspace from low spatial, but high temporal resolution, training data \cite{liang2007spatiotemporal,pedersen2009k,zhao2012image} and follow-up works in which improved self navigated Partially Separable Function (PSF) models were proposed \cite{improved_subspace_estimation_MRI,grasp_pro_MRI}, and (b) methods that enforce the LR constraint in an implicit manner, e.g., via the nuclear or Schatten-$p$ norm regularization with $p < 1$ as in k-t-SLR \cite{lingala2011accelerated} and follow-up work \cite{kt_faster_LR_fmri,kt_faster_LR_fmri_2}. Some of these, such as k-t-SLR \cite{lingala2011accelerated} and PSF-sparse \cite{zhao2012image}, assume both sparsity and LR models on the sequence. 
%

A related line of work models the matrix formed by the MRI sequence as being LR plus sparse (L+S). These methods decompose the dynamic time series as a sum of a LR component modeling smoothly varying time series (e.g. object background, and/or smooth contrast changes as in perfusion MRI), and a sparse component which models the other changes in the image; see \cite{otazo2015low} (L+S-Otazo),  \cite{lin_fessler} (L+S-Lin), and follow-up works, e.g. \cite{tensor_rpca_mri,multiscale_LplusS_lustig}.

Furthermore, motion often breaks down the assumption of low rank in dynamic MRI. There has been extensive work on motion estimation and compensation before imposing the structural assumptions \cite{lingala2014deformation,chen2014motion,tolouee2018nonrigid,feng2016xd}. 

An important challenge with k-t-SLR, L+S-Otazo, L+S-Lin, and most of the above works, is the need for carefully tuning the parameters (regularization parameters and other hyper-parameters associated with the iterative optimization algorithm) for different dynamic MRI applications.
Most published work provides results and code/parameters that work well for only the chosen application (e.g., different set of parameters are provided for cardiac perfusion, and cardiac cine MRI in the open source codes of k-t SLR and L+S-Otazo).
%
A second limitation of the iterative optimization algorithms developed in the above works is that they are slow and memory-inefficient (process the entire matrix $\X$ at each iteration). Both these limitations are exaggerated for the motion compensation methods:  these have even more parameters and are even slower.%

\subsubsection{MRI literature: Deep Learning (DL) methods} There has been much recent work on the use of various DL techniques in the MRI literature. The most common ones are supervised DL reconstruction schemes, e.g., \cite{ke2021learned,ahmed2020dynamic,aggarwal2018modl,arvinte2021deep,herrmann2021feasibility,sandino2021accelerating,zucker2021free,Ghodrati2021temporally,kustner2020cinenet}. These need a large numbers of fully sampled training data points. While such data can be acquired in static imaging applications (e.g., by extending scan times from cooperative volunteers, or compliant patients),  it is not straightforward to acquire sufficient number of fully sampled image sequences for dynamic imaging applications, and definitely not for high time resolution applications, which warrant the need for highly under-sampled acquisitions in the first place. For this reason, a majority of supervised DL models have been used to perform reconstruction frame by frame in dynamic MRI \cite{ke2021learned,ahmed2020dynamic,aggarwal2018modl,arvinte2021deep,herrmann2021feasibility,sandino2021accelerating,zucker2021free}.
This approach does not fully exploit redundancies along the temporal dimension and hence often provides worse reconstructions than sparsity or LR based methods. 
Moreover, DL model learning/training can be very computationally, and hence energy-wise, expensive since each parameter requires retraining the network.  
Finally, the parameters are learned for one specific MRI application, and the same network does not give good results for another application.
Good quality training sequences can be acquired in situations where the motion may be freezed, e.g., in breath held segmented cardiac cine MRI by breathholding, and ECG gating \cite{sandino2021accelerating,zucker2021free,Ghodrati2021temporally,kustner2020cinenet}. These remove the first limitation above, but not the other two. Also, their memory requirement is the biggest limitation.

In recent literature, unsupervised DL based approaches have been proposed and evaluated, these can exploit spatio-temporal redundancies without the need for fully sampled training datasets \cite{ahmed2022dynamic,cole2021fast}. 
However, since these approaches do not use a pre-trained network, but instead train the network on the test/query data, they are orders of magnitude slower compared with both query processing time of supervised DL methods or ours or any of the LR or sparsity based methods. 

\subsubsection{Other related work on online or mini-batch algorithms} Other somewhat related work from the compressive sensing MRI literature that also develops online or minibatch algorithms includes \cite{lassi,on_air}, and follow-up methods.

\subsection{Paper Organization}
We provide the notation and the 3-level approximate-LRcCS  problem formulation in Sec. \ref{problem_set}.
The algorithms are developed in Sec. \ref{all_algos}.  Mini-batch and online subspace tracking approaches are described in Sec. \ref{minibatch_ST_sec}.
 Detailed experimental evaluations and comparisons are provided in Sec. \ref{expts}. Our experimental conclusions and various other issues are discussed in Sec. \ref{discuss}. We conclude in Sec. \ref{conclude}.


\section{Notation and problem formulation}\label{problem_set}

\subsection{Notation and problem setting}
We use $[q]:= \{1,2, \dots, q\}$.
Everywhere, $\|.\|_F$ denotes the Frobenius norm,  $\|.\|$ without a subscript denotes the (induced) $l_2$ norm,  $^\top$ denotes (conjugate) transpose, and $\M^\dagger:= (\M^\top\M)^{-1} \M^\top$.
For a vector $\w$, $|\w|$ computes the magnitude of each on entry of $\w$.
For a scalar $\gamma$, $\indic(\w \le \gamma)$ returns a vector of 1s and 0s of the same size as $\w$ with 1s where $\w(k) \le \gamma$ and zero everywhere else. Here $\w(k)$ is the $k$-th entry of $\w$.
We use $\circ$ to denote the Hadamard product (.* operation in MATLAB). Thus $\z:=\w \circ \indic(|\w| \le \gamma)$ zeroes out entries of $\w$ with magnitude larger than $\gamma$. 
For two $n \times r $ matrices $\Uhat_1, \Uhat_2$ with orthonormal columns, we use $\SEF(\Uhat_1, \Uhat_2): = \|(\I - \Uhat_1 \Uhat_1{}^\top)\Uhat_2\|_F$ as the Subspace Distance (SD) between the subspaces spanned by their columns. 

Let $n$ be the number of pixels in each (unknown) image of the sequence and let $q$ be the total number of images in the sequence.
We denote the vectorized (unknown) image at frame $k$ by $\zstar_k$; this is an $n$-length vector.
We denote the matrix formed by all the $q$ images in the sequence by $\Zstar$. Thus $\Zstar:= [ \zstar_1, \zstar_2, \dots, \zstar_k, \dots, \zstar_q]$ is an $n \times q $ matrix.
The acquired undersampled MRI data/measurements (after some pre-processing) are linear functions of each image. For simplicity of explaining the algorithms (and for comparing the with older theoretical work in this area), we model this linear transformation using a matrix $\A_k$ of size $m_k \times n$.
Thus, our goal is to recover the image sequence matrix $\Zstar$ from
\begin{align}
\y_k : = \A_k \zstar_k , \ k \in [q]
\label{obsmod1}
\end{align}
when  $m_k \ll n$, by making structural assumptions on the matrix $\Zstar$.
\renewcommand{\d}{\bm{d}}
For single-coil dynamic MRI,
\[
 \A_k=\H_k \F
\]
where $\F$ is an  $n \times n$  matrix that models computing the 2D discrete Fourier Transform (DFT) of the vectorized image as a matrix-vector product. The matrix $\H_k $ is a matrix of size $m_k \times n$ with entries being either one or zero. It contains exactly one 1 in each row (corresponding to the observed DFT frequency location converted to 1D coordinates). The mask matrix $\H_k$ is decided by the sampling trajectory (specified in Sec. \ref{expts}).
%

In case of multi-coil dynami MRI with $mc$ coils, there are $mc$ {\cred receive channels }
  with each measuring a subset of Fourier coefficients of a differently weighted version of the cross-section to be imaged.
{\cred
In matrix-vector notation, this can be modeled as follows. Let $\y_{k,j}$ denote the measurements at the $j$-th coil. Then,
\[
\y_k=\begin{bmatrix}
\y_{k,1}\\
\y_{k,2}\\
\vdots\\
\y_{k,mc}
\end{bmatrix}\\=
\underbrace{\begin{bmatrix}
\H_k \F \D_1\\
\H_k \F \D_2\\
\vdots\\
\H_k \F \D_{mc}
\end{bmatrix}}_{\A_k}
\zstar_k
\]
}
where $\D_j = diag(\d_j, j=1,2,\dots,n)$ are $n \times n$ diagonal matrices with diagonal entries (entries of the vector $\d_j$) being the coil sensitivities of the $j$-th coil.
We should point out that $\D_j \x  = \d_j \circ \x$, 
thus $\D_j \x$ is equivalent to weighting the $l$-th pixel $\x_l$ by $(\d_j)_l$.
Each $\y_{k,j}$ is of length $m_k$, thus $\y_k$ is of length $m_k \cdot mc$. 

Let $m=\max_k(m_k)$. We define the $m \times n$ matrix  $\Y= [ (\y_1)_{\mathrm{\tiny{long}}}, (\y_2)_{\mathrm{\tiny{long}}}, \cdots,  (\y_q)_{\mathrm{\tiny{long}}}]$ with $(\y_k)_{\mathrm{\tiny{long}}}$ being the vector $\y_k$ followed by $(m-m_k)$ zeros. Similarly let $(\A_k)_{\mathrm{\tiny{long}}}$ be an $m \times n$ matrix with $(m-m_k)$ rows of zeros at the end. Then, the above model can also be expressed as
{\small
\begin{align}
\Y = \sA(\Xstar):= [(\A_1)_{\mathrm{\tiny{long}}} (\xstar_1), (\A_2)_{\mathrm{\tiny{long}}} (\xstar_2), \dots, (\A_q)_{\mathrm{\tiny{long}}} (\xstar_q)]
\label{Ymat}
\end{align}
}
Similarly $\sA^\top(\Y)$ returns the $n \times q$ matrix 
$\X = \sA^\top(\Y) := [(\A_1)_{\mathrm{\tiny{long}}}^\top (\y_1)_{\mathrm{\tiny{long}}}, (\A_2)_{\mathrm{\tiny{long}}}^\top (\y_2)_{\mathrm{\tiny{long}}}, \dots, (\A_q)_{\mathrm{\tiny{long}}}^\top (\y_q)_{\mathrm{\tiny{long}}}]$.

The above matrix vector model remains valid when the observed samples are available on  Cartesian grid; either they are acquired on a Cartesian grid or are mapped onto a Cartesian grid. 
%
In actual algorithm implementation, all of the above is implemented efficiently using the 2D fast Fourier transform (fft2) function along with appropriate undersampling or use of Hadamard product.
When directly using true radial samples, the main ideas above and in our algorithms given below are still exactly the same, except that fft2 gets replaced by non-uniform FT (NUFT). We use the method of \cite{jeff_nufft} for a fast NUFT.

Writing the model as above makes the ideas easier to follow for readers who are not MRI experts.


\subsection{Hierarchical LR model on image sequence matrix, Z*} \label{approx_lrcs_model}

Most MRI sequences have a certain baseline component that is roughly constant across the entire sequence. Denote this baseline component or ``mean" image by $\zstarbar$.
It can be verified experimentally that this mean image has significantly larger energy compared to the residual image obtained after subtracting  it out. 
Secondly, even after mean subtraction, MR image sequences are only approximately LR, i.e., the residual image obtained after subtracting the mean and the LR components is still not zero, but has a small magnitude compared to the LR component. It is therefore easier to estimate it once the projections of the estimates of the first two components have been subtracted out. Similarly the LR component is easier to estimate once the projections of the mean estimate have been subtracted out.
Thus the following 3-level model is the most appropriate for dynamic MRI: the $k$-th vectorized MR image, $\zstar_k$, satisfies 
\[
\zstar_k = \zstarbar + \xstar_k + \estar_k , \ \forall \ k \in [q],
\]
with the assumption that $\|\estar_k\| \ll \|\xstar_k\| \ll \|\zstarbar\|$, and the $\xstar_k$'s form a rank $r$ matrix $\Xstar:= [\xstar_1, \dots, \xstar_k, \dots, \xstar_q]$ with $r \ll \min(n,q)$.
Here $\estar_k$ is the unstructured residual signal component, which we refer to as the modeling error. We will consider two models on $\estar_k$. The first does not assume any structure on $\estar_k$s except assuming that their magnitude is small. The second assumes that $\estar_k$s are small magnitude, and sparse in the temporal Fourier domain (rows of the matrix $\Estar$ are Fourier sparse).%

The first model can be interpreted as a 3-level LR model: the first level is a special case of the LR model with rank $1$: the ``mean image'' matrix, $\zstarbar \bm{1}^\top$ (where $\bm{1}$ is a vector of $q$ 1's) has $rank=1$; the second level is matrix $\Xstar$ which has $rank=r$; and the third level is the $rank=\min(n,q)$ matrix $\Estar$. Our assumption implies $\|\zstarbar \bm{1}^\top\|_F \gg \|\Xstar\|_F \gg \|\Estar\|_F$.

\section{AltGDmin-MRI algorithms}\label{all_algos}

\subsection{AltGDmin-MRI overall idea}
We develop a 3-level hierarchical algorithm that first recovers $\zstarbar$, then the $\xstar_k$s, and then $\estar_k$s. 
%
Under the modeling assumption that $\|\zstarbar\| \gg \|\xstar_k\| \gg |\estar_k\|$, the recovery of $\zstarbar$  becomes the following least squares (LS) problem:
\[
\min_{\tilde\zbar} \sum_{k=1}^q \|\y_k - \A_k \tilde\zbar\|^2.
\label{minx}
\]
Denote its solution by $\zbar$. 
Next, we estimate the rank-$r$ matrix $\Xstar$ (and the rank $r$ itself) from the measurement residuals,
\[
\ty_k:= \y_k - \A_k \zbar, \ k \in [q]
\label{yres1}
\]
by using an automated version of altGDmin for LRcCS \cite{lrpr_gdmin} applied to $\ty_k$s. This is described below in Sec. \ref{agm}.
Denote its output by $\Xhat$.
The last step, which we refer to as Modeling Error Correction (MEC), involves estimating the modeling error $\estar_k$ from the new measurement residuals
\[
\tilde{\ty}_k:= \y_k - \A_k\zbar - \A_k \xhat_k, \ k \in [q]
\label{yres2}
\]
Depending on which of the two models is assumed on $\estar_k$, the steps to estimate it are different. We describe them in Sec. \ref{mec}. Denote the output of either step by $\Emat$.

The final output is $\Z:= [\z_1, \z_2, \dots, \z_q]$ with $\z_k = \zbar + \xhat_k +  \e_k$. We summarize these steps in Algorithm \ref{gdmin_framework}.

\begin{algorithm}[t]
\caption{\sl{altGDmin-MRI. CGLS is the code from \cite{cgls_code}.}}
\label{gdmin_framework}
\ben

\item  Solve $\min_{\bar\z} \sum_{k=1}^q \|\y_k - \A_k \bar\z\|^2$ using CGLS with tolerance $10^{-3}$ and maximum number of iterations 10.
Denote the solution  by $\zbar$.

\item
\ben
\item For each $k \in q$, compute $\ty_k:= \y_k - \A_k \zbar$.

\item Run Algorithm \ref{gdmin_prac} (auto-altGDmin) with $\ty_k, \A_k$ as its inputs. Its output is $\X$.
\een

\item
\ben
\item For each $k \in q$, compute $\tilde{\ty}_k:= \y_k - \A_k \zbar - \A_k \xhat_k$.

\item Run step 1 (altGDmin-MRI1)  OR step 2  (altGDmin-MRI2) of  Algorithm \ref{gdmin_mec} 
 The output of either is the matrix $\Emat$.
\een


\een
Output $\Z:= [\z_1, \z_2, \dots, \z_q]$ with $\z_k = \zbar + \xhat_k +  \e_k$.
\end{algorithm}

\begin{algorithm}[t]
\caption{\sl{auto-altGDmin: altGDmin with automated parameter setting. Let $\M^\dagger:= (\M^\top\M)^{-1} \M^\top$.}}
\label{gdmin_prac}
\begin{algorithmic}[1]
   \State {\bfseries Input:} $\ty_k, \A_k, k \in [q]$.
   \State {\bfseries \color{blue}  Parameter setting is specified in {blue}}.
\State {\bf Initialization:}
\State Compute $\gamma= 36 \sum_\ik|\ty_\ik|^2 / mq$, $\ty_{k,\tiny{tnc}} = \ty_k \circ \indic\{|\ty_k| \le \sqrt{\gamma} \}$, $\mbar = \sum_{k=1}^q m_k/q$, and compute 
{\small
\begin{align*}
\X_{0} & :=  &   \left[  \frac{1}{\sqrt{m_1 \mbar}} (\A_1^\top \ty_{1,\tiny{tnc}}),  ..., \frac{1}{\sqrt{m_k \mbar}} (\A_k^\top \ty_{k,\tiny{tnc}}), ..., \right.  \\
&& \left. \frac{1}{\sqrt{m_q \mbar}}(\A_q^\top \ty_{q,\tiny{tnc}}) \right]
\end{align*}
}

\State {\color{blue} Let $\sigma_j = \sigma_j(\X_0)$. Set $\rhat$ as the smallest integer for which
$$\sum_{j=1}^r \sigma_j^2 \ge (b/100) \cdot \sum_{j=1}^{\min(n,q,mc \min_k m_k )/10} \sigma_j^2, \ b=85.$$
}

\State  Set $\Uhat_0 \leftarrow $ top $\rhat$ left singular vectors of $\X_0$

\State {\bf GDmin iterations:} {\color{blue} Set $T_{max} = 70$}
  \For{$t=1$ {\bfseries to} $T_{max}$}

   \State  Let $\Uhat \leftarrow \Uhat_{t-1}$.

 \State Update $\Bhat$: For all $k \in [q]$,  $\bhat_k   \leftarrow  (\A_k  \U)^\dagger \ty_k $.
   \State Update $\Xhat$:  For all $k \in [q]$,  $\xhat_k   \leftarrow \Uhat \bhat_k $

  \State Gradient compute:
  $$\nabla_U f (\Uhat, \Bhat) \leftarrow \sum_{k=1}^q \A_k^\top (\A_k \Uhat \bhat_k - \ty_k) \bhat_k^\top$$

  \State {\color{blue} If $t=1$, set $\eta = 0.14/ \|\nabla_U f (\Uhat, \Bhat)\|$.}

  \State   GD step for $\Uhat$: $\ds \Uhat^+   \leftarrow \Uhat - \eta \nabla_U f(\Uhat, \Bhat)$
   \State   Projection for $\Uhat$: $\Uhat^+ \qreq \U^+ \R^+$. 
 Set $\Uhat_t \leftarrow \U^+$.

\State {\color{blue} EXIT loop if $\SE(\U, \U^+)/\sqrt{\rhat} < \epsilon_{exit}=0.01$
} 
\EndFor
\State      {\bfseries Output:} $\Xhat:= [\xhat_1, \xhat_2, \dots, \xhat_q]$.
\end{algorithmic}
\end{algorithm}

\subsection{Automated AltGDmin}\label{agm}

Recall that $\ty_k$'s are the measurement residuals after subtracting the projections of the estimated mean. Our next goal is to estimate a rank $r$ matrix $\X$ that minimizes
\[
\tilde{f}(\X): = \sum_{k=1}^q \|\ty_k - \A_k \x_k\|^2
\]
\subsubsection{Motivation for a novel GD-based algorithm}
We would like a GD based solution since those are known to be much faster than both AltMin and convex relaxation methods \cite{rmc_gd,lrpr_gdmin}.
As explained in detail in  \cite{lrpr_gdmin}, neither of the two commonly used GD approaches from LR recovery literature, and LR matrix completion (LRMC) in particular, -- projected GD on $\X$ (projGD-$\X$) \cite{rmc_gd} or alternating GD with a norm balancing term (altGDnormbal)  \cite{rpca_gd,lafferty_lrmc} -- provably works for the LRcCS measurement model. The reason is that, in both cases, the estimates of the columns $\xstar_k$ are too coupled.
%
Moreover, even for LRMC for which these approaches do work, projGD-$\X$ is memory-intensive: it requires memory of order $nq$; while altGDnormbal is slow: it needs a GD step size that is $1/r$ or smaller \cite{rpca_gd,lafferty_lrmc}, making it $r$-times slower than GD with a constant step size.
%
{\cred
{\em The following modification, that we dub {\em altGDmin}, is as fast per-iteration as projGD-$\X$, as memory-efficient as altGDnormbal, and yet its estimate are only mildly coupled (they are uncoupled given an estimate of the column span of $\Xstar$).} Because of this, AltGDmin is amenable to analysis that helps show that, in the random Gaussian measurement setting, the algorithm converges fast -- the required number of iterations to achieve $\epsilon$ accuracy grows as $\log(1/\epsilon)$ -- while needing a small number of samples \cite{lrpr_gdmin}.%
}

\subsubsection{AltGDmin algorithm}
Rewrite the unknown matrix $\X$ as $\X= \U \B$, where $\U$ is $n \times r$ and $\B$ is $r \times q$, and consider
$$f(\U,\B):= \tilde{f}(\U \B) =  \sum_{k=1}^q \|\ty_k - \A_k \U \b_k\|^2.$$
AlGDmin involves iterating over the following two steps after starting with a carefully designed initialization for $\U$.

\ben
\item For each new estimate of $\U$, we solve for $\B$ by minimizing over it while keeping $\U$ fixed at its current value.
Because our measurements are column-wise decoupled ($\ty_k$ does not depend on any other image except the $k$-th one), the minimization step gets decoupled for the different columns of $\B$, i.e.,
\[
\min_{\B} f(\U,\B) = \sum_{k=1}^q \min_{\b_k} \|\ty_k - \A_k \U \b_k\|^2.
\]
This problem is now a very quick column-wise least squares (LS) problem with $\b_k$ being an $r$-length vector and $\A_k \U$ being an $m_k \times r$ matrix. Thus, the complexity is only $m_k r^2 \cdot q$ (for solving $q$ individual LS problems) plus the cost of computing $\A_k \U$ for all $k \in [q]$. 
The total complexity for this step is thus similar to that of one GD step for $\U$.
\item  We use projected GD for updating $\U$: one GD step w.r.t. $\U$ followed by projecting onto the space of orthonormal matrices (by using QR decomposition).

\een


Finally, since $f(\U,\B)$ is not convex in the unknowns $\{\U, \B\}$, the above algorithm needs a careful initialization of one of them. 
Using the standard approach from LR recovery literature, we can initialize $\U$ by computing the top $r$ left singular vectors of the matrix
\bea
\X_0 =    \left[ \frac{1}{m} \A_1^\top \ty_{1},  \dots,  \frac{1}{m} \A_k^\top \ty_{k}, \dots,  \frac{1}{m} \A_q^\top \ty_{q} \right]
 \nonumber
\eea

When the measurement matrices $\A_k$ are random Gaussian, a truncation/thresholding step, that zeros out entries of $\ty_k$ with magnitude much larger than the root mean squared value $\sqrt{\sum_\ik \ty_\ik^2/mq}$, is required on each $\ty_k$ before computing the above matrix. This step helps to filter out the very large measurements (those whose $\ty_\ik^2$ is much larger than the expected value) and helps ensure that the new matrix has entries which are sub-Gaussian\footnote{In our proofs, this allows us to use the sub-Gaussian Hoeffding inequality  \cite{versh_book} to get our desired bound the initialization error.}.
In the MRI setting, since the matrices $\A_k$ are subsampled Fourier, if the measurements are indeed noise-free as assumed in \eqref{obsmod1} and the matrix satisfies the incoherence assumption given earlier, then the above is not needed. The reason is, in this case, $\|\a_\ik\|^2 = 1$ \footnote{or some constant that is the same for all $i,k$ and depends on how the Fourier matrix is normalized} and so, $\ty_\ik^2 \le \max_k \|\zstar_k - \zbar\|^2$ for all $i,k$. However, in practice, there have be either measurement or image outliers: in some acquisitions, there may be occasional large noise or, certain images $\zstar_k$ may be outliers, e.g., this would happen if the subject took a deep breath during acquisition. The truncation step helps filter out such measurements. 
If there are no outliers or large noise, then it does nothing and hence it does not worsen performance either.
 Another minor change is needed: since $m_k \neq m_1$ (time varying number of measurements), in order to prove a guarantee similar to our result from \cite{lrpr_gdmin}, one needs to replace the $1/m$ factor by $1/\sqrt{m_k \mbar}$ where $\mbar=\sum_{k=1}^q m_k/ q$. We specify $\X_0$ with these modifications in line 5 of Algorithm \ref{gdmin_prac}.




\subsubsection{AltGDmin parameter setting} \label{param_set} 
The parameters for AltGDmin are the rank $r$, the GD step size $\eta$, and the maximum number of iterations $T$ along with a stopping criterion to exit the loop sooner if the estimates do not change much. 

For approximately LR matrices, there is no one correct choice of $r$. We use the following constraints to find a good approach. We need our choice of rank, $\rhat$, to be sufficiently small compared to $\min(n,q)$ for the algorithm to take advantage of the LR assumption. Moreover, for the  LS step for updating $\bhat_k$'s to work well (for its error to be small), we also need it to be small compared with $mc \min_k m_k$. Based on just these constraints, one can set $\rhat = \min(n,q,mc \min_k m_k)/10$. Or, one can compute the ``$b\%$ energy threshold'' of the first $\min(n,q,mc \min_k m_k)/10$ singular values, 
 i.e., compute $\rhat$ as the smallest value of $r$ for which
\[
\sum_{j=1}^\rhat \sigma_j(\X_0)^2 \ge (b/100) \cdot \sum_{j=1}^{\min(n,q,mc \min_k m_k )/10} \sigma_j(\X_0)^2.
\]
for a $b \le 100$. Here $\sigma_j(\X_0)$ is its $j$-th singular value.
We use this latter approach with $b = 85$. We have experimented with other values as well in the 80-95\% range, and the algorithm is not very sensitive to this choice.


We set the GD step size $\eta = 0.14 / \|\nabla_U f(\Uhat_0 , \Bhat_0)\|$ where $\Uhat_0, \Bhat_0$ are the initial estimates. Assuming that the gradient norm decreases over iterations, this implies that $\eta \cdot \|\nabla_U f(\Uhat_t , \Bhat_t)\| \le 0.14 < 1$ always. Since $\|\Uhat_t\| = 1$ (due to the QR decomposition step), this ensures that a GD step is never too big.
%
To decide $T$ (maximum number of iterations), we stop the GD loop when $\SE(\Uhat_{t-1}, \Uhat_t) < \epsilon_{exit} \sqrt{r}$  while setting $T_{max} = 70$ so that no more than 70 iterations are run. We set $\epsilon_{exit}=0.01$. 

We summarize the complete algorithm, with the above parameter settings, in Algorithm \ref{gdmin_prac}.


As suggested in \cite{lrpr_gdmin}, one can also set $\eta$ as $\eta = c/(m\|\Uhat_0  \Bhat_0\|^2)$ with a $c < 1$. This is a conservative approach that is needed for proving guarantees which are only sufficient conditions and will lead to slower convergence. 

{\cred
\subsubsection{AltGDmin guarantee}
Theorem 2.1 of \cite{lrpr_gdmin} proved the following for AltGDmin with parameters as specified there.
\begin{theorem}
Suppose that $\Zstar = \Xstar$, i.e., it is a matrix with rank $r$, and $\ty_k = \y_k$. Suppose also that each $\A_k$ is $m \times n$ and i.i.d. random Gaussian.
Assume that $\max_k \|\xstar_k\|^2 \le \mu^2 \|\Xstar\|^2 /q $ for a constant $\mu$ (incoherence parameter) that is only a little larger than one. Let $\xhat_k$ denote the AltGDmin estimates after $T = C \kappa^2 \log(1/\epsilon)$ iterations. If
$mq \ge C\kappa^4 \mu^2 (n+q)r^2 \log(1/\epsilon)$, if the algorithm parameters are set as described in  \cite[Theorem 2.1]{lrpr_gdmin},
then, with probability at least $1 - n^{-10}$, $||\xstar_k - \xhat_k|| \le \epsilon ||\xstar_k||$ for all $k=1,2,\dots,q$.
\end{theorem}
In the above result $\kappa$ is the ratio of the first to the $r$-th singular value of $\Xstar$. Treating $\kappa,\mu$ as numerical constants, this is equivalent to requiring that $\max_k \|\xstar_k\|^2 \le C \|\Xstar\|_F^2/q$ for a constant $C$. In practice, this means that, the different (vectorized) images $\xstar_k$ in the sequence have similar enough energy so that the maximum energy of any one of them is not much larger than its average, $\|\Xstar\|_F^2/q$. This fact is very valid for MRI datasets.
}

\subsection{Model error correction (MEC): two models and algorithms} \label{mec} 

\subsubsection{MEC using Model 1 (no structure on $\Estar$): altGDmin-MRI1}
Recall that this model assumes no structure on $\estar_k$ except that it has small magnitude. We thus recover each $\e_k$ individually by solving
\[
\min_{\e} \|\tilde{\ty}_k - \A_k \e\|^2
\]
for each $k$, while imposing the assumption that $\|\e\|^2$ is small. An indirect way to enforce this, while also getting a fast algorithm, is to start with a zero initialization and run only a few iterations of GD to solve the above minimization problem.



\subsubsection{Parameter setting for MEC-1}
We use the Stanford Conjugate Gradient LS (CGLS) code \cite{cgls_code} for solving the above minimization.
We used this code with tolerance of $10^{-36}$ and maximum number of iterations 3. 

This is summarized in Algorithm \ref{gdmin_mec}.

\subsubsection{MEC using Model 2 (Temporal Fourier sparse $\Estar$): altGDmin-MRI-2} \label{mec2}
Our second model assumes Fourier sparsity of the modeling error along the time axis. 
To be precise,  we are assuming that 
\[
\Sstar:= \sF(\Estar)
\]
is a row sparse matrix (matrix whose rows are sparse vectors). Here, the operator $\sF$ computes the 1D DFT of each row of its argument.
We thus have the following model on the images' matrix $\Zstar$:
\[
\Zstar = (\zstarbar \bm{1}^\top) + \Xstar + \Estar, \ \ \Estar := \sF^{-1} (\Sstar)
\]
with $\|\Estar\|_F \ll \|\Xstar\|_F \ll \sqrt{q} \|\zstarbar\|$ and $\Sstar$ being row sparse.

%

To estimate $\Estar$ under this model, we use the Iterative Soft Thresholding Algorithm (ISTA) for sparse recovery \cite{ista_ref} which was also used in \cite{otazo2015low}. For recovering an unknown sparse $\s$ from $\y:= \A \s$, this starts with a zero initialization, $\s=0$, and runs the iterations: $\s \leftarrow \scriptsize{\mathrm{SThr}}_{\omega}(\s + \A^\top(y - \A \s))$. Here $\scriptsize{\mathrm{SThr}}_{\omega}(\s)$ is the Soft-Thresholding operator; it zeroes out entries of $\s$ that are smaller than $\omega$ while shrinking the larger magnitude entries towards zero by $\omega$, i.e. $[\scriptsize{\mathrm{SThr}}_{\omega}(\s)]_i = sign(\s_i) (|\s_i| - \omega)$ if $|\s_i|> \omega$ and $[\scriptsize{\mathrm{SThr}}_{\omega}(\s)]_i =0$ otherwise.

For our model, this translates to the following iteration. Compute the residual $\tilde{\tilde{\Y}}:= \Y - \sA( \zbar \bm{1}^\top) - \sA(\X)$.
Update $\Emat$ by running the following iteration starting with $\Emat=0$:
\[
\Emat \leftarrow \sF^{-1} ( \scriptsize{\mathrm{SThr}}_{\omega}(  \sF( \Emat +  \sA^\top ( \tilde{\tilde{\Y}} - \sA(\Emat) ) ) ) )
\]
This is summarized in Algorithm \ref{gdmin_mec}.

The temporal Fourier sparsity model has been used for imposing the L+S assumption for dynamic MRI in \cite{otazo2015low}, \cite{lin_fessler}, and follow-up works. We should clarify that, in  this work, we are not imposing the L+S model, instead we are assuming a 3-level hierarchical model, with sparsity being used to model the residual in the third level.
The assumption $\|\Estar\|_F \ll \|\Xstar\|_F$ makes our model different from the regular L+S model which assumes $\Zstar =\Xstar + \Estar$ with no assumption on one of them being smaller in magnitude than the other.
From our experiments in Sec. \ref{expts}, in an average sense, our algorithm, altGDmin-MRI2 that uses our model, gives better reconstructions (both in terms of error and visually). However, this may be either because the assumed models are different or because the reconstruction algorithms are very different too: we use a GD-based algorithm, while \cite{otazo2015low}, \cite{lin_fessler}, use different algorithms to solve the convex relaxation of the L+S model. This issue will be explored in more detail in future work where we plan to also develop and evaluate a GD-based algorithm under the L+S assumption.

\subsubsection{Parameter setting for MEC-2} We  used soft thresholding with threshold as given in Algorithm \ref{gdmin_mec}. 


\subsection{Implementation}
We write things as above only for ease of explanation. In our implementation, we never use matrix-vector multiplication for computing $\A_k \x$ or $\A_k^\top \y$,  since that is much more memory intensive and much slower than using Fast FT (FFT). Also, we use various MATLAB features and linear algebra tricks to remove ``for'' loops wherever possible. Code is provided at the link given in Sec. \ref{expts}.
%
%


\begin{algorithm}[t]
\caption{\sl{altGDmin-MRI1 or altGDmin-MRI2: }}
\label{gdmin_mec}

\begin{itemize}
\item  {\bf Unstructured MEC (altGDmin-MRI1)}
\\ Run the following
\begin{itemize}
\item For each $k \in [q]$, run 3 iterations of CGLS to solve $\min_{\e} \|\tilde{\ty}_k - \A_k \e\|^2$. Denote the output by $\e_k$.
\end{itemize}
or
\item {\bf Sparse MEC (altGDmin-MRI2)}
\\ Run the following ISTA algorithm
\begin{itemize}
\item   $\Emat_0 = \bm{0}$, 
 $\tau = 0$. Repeat the following steps

\begin{itemize}
\item $\Mmat_\tau \leftarrow \sF( \Emat_\tau+\sA^\top(\tilde{\tilde{\Y}} - \sA(\Emat_\tau)) )$

\item If $\tau=0$, set $\omega =  0.001 \cdot \|\Mmat_0\|_{\max}$ where $\|.\|_{\max}$ is the maximum magnitude entry of the matrix.

\item $\tau \leftarrow \tau+1$

\item
$
\Emat_\tau \leftarrow \sF^{-1} ( \scriptsize{\mathrm{SThr}}_{\omega}(  \Mmat_\tau  ) )
$

\end{itemize}
Until $\tau= 10$ or $\frac{\|\Mmat_\tau - \Mmat_{\tau-1}\|_F}{\|\Mmat_{\tau-1}\|_F} <  0.0025$ 

\end{itemize}

\end{itemize}
Output $\Emat$.
\end{algorithm}

\input{revise_dynmri_subtrack}

\input{revise2_dynmri_lrcs_figstables}

\input{revise2_dynmri_lrcs_expts}

\input{revise2_dynmri_lrcs_discussion}

\bibliographystyle{IEEEtran}
\bibliography{../bib/tipnewpfmt_kfcsfullpap,./refs,./refs_Silpa}

\appendices
\input{revise_dynmri_lrcs_appendix}

\end{document}

%% file: revise_dynmri_subtrack.tex
\section{AltGDMin based Subspace Tracking} \label{minibatch_ST_sec}
The algorithms discussed so far are batch methods, i.e., they require waiting for all the measurements to be taken. This means that they cannot be used in applications that require near real-time reconstructions (explained in Sec. \ref{contrib}). 


\subsection{Mini-batch Subspace Tracking}
Consider the pseudo-real-time setting in which the algorithm processes each new mini-batch of the data (here a set of $\alpha$ consecutive $\y_k$s) as soon as it arrives. Thus, instead of waiting for all $q$ measurements $\y_k$ to be obtained, it only waits for a new set of $\alpha < q$ measurements before processing them. 
For algorithms that use the LR assumption on the data, such an algorithm can be referred to as a {\em Subspace Tracking} solution since it is implicitly assuming that consecutive mini-batches of data lie  close to the same or slightly different $r$-dimensional subspaces. 
One can utilize this ``slow subspace change'' assumption in the following fashion. For the first mini-batch, use the altgdmin-MRI algorithm with $q$ replaced by $\alpha$. For later mini-batches, use altgdmin-MRI with two changes.
(1) Use the final estimated $\Uhat$ from the previous mini-batch, denoted $\Uhat^{(j-1)}$, as the initialization for the current one. This means that we replace lines 2-6 of Algorithm \ref{gdmin_prac} by $\Uhat_0 \leftarrow \Uhat^{(j-1)}$. (2) Second, reduce the maximum number of iterations for the $j$-th minibatch, denoted $T_{\max,j}$, to a much lower value for $j > 1$ than for $j=1$.

The complete algorithm is summarized in Algorithm \ref{gdmin_ST}. We use $T_{max,1}=70$ and $T_{max,j}=5$ for $j\ge 2$.
Depending on the MEC step being used, we refer to the resulting algorithms as altGDminMRI-ST1 and altGDminMRI-ST2. 




\begin{algorithm}[t!]
\caption{\sl{altGDminMRI-ST1 and altGDminMRI-ST2: Mini-batch Subspace Tracking with MEC model 1 (ST1) or model 2 (ST2)}}
\label{gdmin_ST}
\ben
\item Let $j=1$. Run Algorithm \ref{gdmin_framework} on first mini-batch of $\alpha_1$ $\y_k$s. Thus $q \equiv \alpha_1$ and we let $T_{\max,1}=70$.
Denote its final subspace estimate by $\Uhat^{(1)}$.

\item For each $j > 1$ do 
\ben

\item Run Algorithm \ref{gdmin_framework} with the following two changes: (i) replace the Initialization step of altGDmin (Algorithm \ref{gdmin_prac}) by $\Uhat_0 \leftarrow \Uhat^{(j-1)}$; and (ii) set $T_{\max,j}=5$. 
    
    Denote its final subspace estimate by $\Uhat^{(j)}$.
\een
End For
\een

\end{algorithm}

\Subsection{Online Subspace Tracking}
In certain other applications,  after an initial short delay, a true real-time  (fully online) algorithm is needed.  This means that, each time a new $\y_k$ is obtained, it should return a new estimate $\xhat_k$. To obtain such an algorithm we eliminate the mean computation step and the $\U$ update steps in online mode. Both these are computed only for the first mini-batch and used at all later times. 
Suppose the first mini-batch consists of $\alpha_1$ frames. For all times $k > \alpha_1$, we use the estimated mean $\zbar$ and the estimated $\Uhat$ from the first mini-batch. For $k > \alpha_1$, for each new $\y_k$, we only update $\b_k$ and $\e_k$ and only using MEC model 1 (unstructured $\estar_k$). 
%
%
 We summarize this algorithm, altGDminMRI-onlineST, in Algorithm \ref{gdmin_onlineST}. 


\begin{algorithm}[t!]
\caption{\sl{altGDminMRI-onlineST: Online Subspace Tracking}}
\label{gdmin_onlineST}
\ben
\item For the first mini-batch of $\alpha$ frames, run Algorithm \ref{gdmin_framework}. Denote the computed mean image by $\zbar^{(1)}$ and the final subspace estimate by $\Uhat^{(1)}$.

\item For $k > \alpha+1$ do:
\ben
\item Compute $\ty_k = \y_k - \A_k \zbar^{(1)}$
\item  Let $\Uhat =  \Uhat^{(1)}$.
\item Compute $\bhat_k   \leftarrow  (\A_k  \U)^\dagger \ty_k $.
\item Compute $\tilde{\ty}_k = \ty_k  - \A_k \Uhat \bhat_k$
\item Compute $\e_k$  by running 3 iterations of CGLS to solve $\min_{\e} \|\tilde{\ty}_k - \A_k \e\|^2$. Denote the output by $\e_k$.
\item Output $\xhat_k = \zbar_{(1)} + \Uhat \bhat_k + \e_k$
\een
End For

\een

\end{algorithm}

%
%
%

%
%

%% file: revise2_dynmri_lrcs_figstables.tex
\begin{table}
\begin{center}
{
\begin{tabular}{|c|c|c|c|c|}
\hline
 & altGDmin &   altGDmin-mean  &  AltMin & MixedNorm  \\
\hline
&&&& \\
Gauss.  &0.010 (0.15)&  0.009 (0.18) &  0.113 (13) &0.029 (42)\\
\hline
&&&& \\
Four. &0.331 (0.50)& 0.002 (0.55)& 0.025 (86) & 1.0 (145)\\
\hline
\end{tabular}
}
\vsm
\end{center}
\caption{\sl\small{
We report {\bf Error (Reconstruction Time in seconds)}. Here Error is the Monte Carlo average of $\|\Xhat - \Xstar\|_F^2 / \|\Xstar\|_F^2$ over 50 realizations.
Comparisons on a 30 x 30 x 90 image piece of the PINCAT sequence ($n= 900, q=50$) using $m =n/10$ random Gaussian (Gaussian) or random Fourier  (Four.) measurements. 
}}
\vsm
\label{theory_algos1}
\end{table}

\begin{table*}[t]
\begin{center}
\begin{tabular}{|l|l|l|l|l|l|l|l|}
\hline
Dataset  & kt-SLR & L+S-Otazo & L+S-Lin &  altGDmin-MRI1 & altGDmin-MRI2 & altGDminMRI- & PSF-sparse\\
&&&&&&ST2, $\alpha$=64,100&\\
\hline
\multicolumn{6}{|l|}{Cartesian} 
\\
\hline
CardPerf-R8 &   0.6398 (420.71)&   0.0110 (16.93)&   0.0292 ( 5.99) &   0.0201 (12.01)&   0.0194 (14.74) &&\\
\hline
CardCine-R6 &   0.1377 (614.53)&   0.0054 (24.57)&   0.0069 (15.51)&     0.0058 (37.55)&   0.0055 (46.08) &&\\
\hline
\multicolumn{6}{|l|}{Pseudo-radial}  \\
\hline
Brain(4) &   0.0093 (127.92)&   0.0167 ( 5.65)&   0.0173 ( 2.53) &   0.0125 ( 3.70)&   0.0121 ( 4.86)&& \\
Brain(8) &   0.0034 (102.33)&   0.0086 ( 4.15)&   0.0095 ( 2.48) &   0.0054 ( 3.87)&   0.0051 ( 5.11)&& \\
Brain(16) &   0.0014 (75.57)&   0.0049 ( 2.81)&   0.0062 ( 2.44) &   0.0027 ( 3.84)&   0.0024 ( 4.98) &&\\
\hline
%
Speech(4) &   0.1543 (5234.83)&   0.1991 (537.91)&   0.2545 (304.50) &   0.1416 (131.87)&  0.1395 (153.43) &  0.1174 (86.57) &\\
Speech(8) &   0.0593 (5261.49)&   0.1107 (491.11)&   0.1284 (306.16) &   0.0991 (152.35)& 0.0952 (176.35)  &   0.0873 (87.64)&  \\
Speech(16) &   0.0203 (5288.61)&   0.0550 (426.10)&   0.0557 (304.45) &   0.0580 (236.85)& 0.0540 (261.39) &  0.0542 (100.75) &   \\
\hline
UnCardPerf(4) &   0.0894 (4150.72)&   0.0910 (189.88)&   0.1424 (50.68)&    0.0695 (70.97)&   0.0684 (92.31)&   & \\ 
UnCardPerf(8) &   0.0442 (3472.96)&   0.0591 (120.55)&   0.0632 (50.44)&    0.0470 (67.79)&   0.0451 (90.66)& 0.0531 (81.48) & \\
UnCardPerf(16) &   0.0206 (2873.30)&   0.0370 (88.44)&   0.0329 (50.47)&    0.0298 (69.08)&   0.0275 (90.16)&0.0328 (70.77) & \\
\hline
CardOCMR16(4) &   0.0362 (227.92)&   0.0293 (10.73)&   0.0515 ( 2.75) &   0.0092 (10.52)&   0.0092 (15.42) &&0.3803 (2.39)\\
CardOCMR16(8) &   0.0045 (225.98)&   0.0064 ( 8.37)&   0.0101 ( 2.73)&      0.0033 ( 7.26)&   0.0033 ( 8.64) &&0.1200 (7.91)\\
CardOCMR16(16) &   0.0015 (162.12)&   0.0035 ( 4.64)&   0.0030 ( 2.73)&      0.0015 ( 5.39)&   0.0014 ( 6.69)&& 0.0020 (5.24) \\
\hline
CardOCMR19(4) &   0.0216 (399.70)&   0.0251 (18.70)&   0.0698 ( 5.06) &   0.0095 (11.58)&   0.0094 (14.10)& &\\
CardOCMR19(8) &   0.0043 (409.01)&   0.0092 (13.05)&   0.0149 ( 5.06)&      0.0051 (10.49)&   0.0050 (12.91) & &\\
CardOCMR19(16) &   0.0020 (269.01)&   0.0052 ( 7.20)&   0.0044 ( 5.05)&     0.0032 ( 9.47)&   0.0030 (12.00)& &\\
\hline
PINCAT(4) &   0.0445 (34.85)&   0.0381 (8.04) &   0.1054 (2.26)&     0.0278 (1.63)&   0.0278 (1.77)  & &\\
PINCAT(8) &   0.0216 (31.54)&   0.0162 (3.91) &   0.0208 (2.22)&       0.0166 (1.36)&   0.0166 (1.31)  & &\\
PINCAT(16) &   0.0095 (23.31)&   0.0065 (2.70) &   0.0047 (2.25)&       0.0097 (1.08)&   0.0097 (1.13) & &\\
\hline
av-Err (av-Time) & 0.0663 (1470.3) & 0.0369 (99.3) & 0.0515 (56.3)  & 0.0289 (42.4) & 0.0280 (50.7)& &\\
\hline
\end{tabular}
\end{center}
\vsm
\caption{\sl\small{Table format is {\bf Error (Recon time in seconds)}. The last row shows {\bf average-Error (average-Reconstruction time in seconds)} over all 20 rows of results. For PSF-Sparse, we generated data using the k-t sampling scheme from their paper \cite{zhao2012image} while ensuring same total number of samples as the rest of the compared methods.
}}
\label{table_multicoil}
\vsm
\end{table*}

\begin{table}[h]
\small
\begin{center}
{
\begin{tabular}{|c|c|c|c|}
\hline
$\alpha$ & 	altGDminMRI-  & 	altGDminMRI- & altGDminMRI- \\
&ST1 &ST2 &onlineST\\
\hline

\multicolumn{3}{|c|}{Cardiac 16 radial lines}  \\
\hline
100 &   0.0355 (50.20) &   0.0328 (70.77) &   0.0782 (37.61)\\
50 &   0.0349 (45.29) &   0.0323 (67.47) &   0.0958 (27.35)\\
\hline
\multicolumn{3}{|c|}{Cardiac 8 radial lines}  \\
\hline
100 &   0.0556 (59.51) &   0.0531 (81.48) &   0.1038 (48.05)\\
50 &   0.0579 (52.17) &   0.0555 (75.73) &   0.1271 (36.61)\\
\hline
\multicolumn{3}{|c|}{Cardiac 4 radial lines}  \\
\hline
100 &   0.0705 (135.54) &   0.0695 (164.97) &   0.1273 (138.44)\\
50 &   0.0750 (87.31) &   0.0737 (109.71) &   0.1586 (70.05)\\
\hline
\multicolumn{3}{|c|}{Speech 16 radial lines}  \\
\hline
1024 &   0.0575 (192.93) &   0.0534 (219.95) &   0.0895 (136.76)\\
512 &   0.0578 (148.53) &   0.0534 (175.08) &   0.1139 (94.19)\\
256 &   0.0584 (109.91) &   0.0538 (136.52) &   0.1318 (63.22)\\
128 &   0.0584 (86.09) &   0.0538 (113.76) &   0.1468 (46.94)\\
64 &   0.0589 (74.05) &   0.0542 (100.75) &   0.1613 (40.43)\\
32 &   0.0604 (67.24) &   0.0554 (93.57) &   0.1771 (37.45)\\
\hline
\multicolumn{3}{|c|}{Speech 8 radial lines}  \\
\hline
1024 &   0.0888 (189.69) &   0.0858 (212.63) &   0.1260 (143.03)\\
512 &   0.0882 (124.93) &   0.0850 (154.88) &   0.1525 (89.61)\\
256 &   0.0885 (93.39) &   0.0851 (119.25) &   0.1735 (59.94)\\
128 &   0.0889 (74.42) &   0.0853 (98.52) &   0.1933 (45.93)\\
64 &   0.0911 (66.05) &   0.0873 (87.64) &   0.2115 (40.23)\\
32 &   0.1007 (61.85) &   0.0960 (83.86) &   0.2292 (36.87)\\
\hline
\multicolumn{3}{|c|}{Speech 4 radial lines}  \\
\hline
1024 &   0.1202 (576.38) &   0.1191 (594.78) &   0.1642 (537.66)\\
512 &   0.1257 (212.70) &   0.1242 (241.69) &   0.2023 (184.72)\\
256 &   0.1201 (93.79) &   0.1183 (119.46) &   0.2226 (63.93)\\
128 &   0.1177 (71.25) &   0.1157 (96.15) &   0.2490 (46.26)\\
64 &   0.1195 (65.27) &   0.1174 (86.57) &   0.2715 (41.74)\\
32 &   0.1321 (62.40) &   0.1293 (83.83) &   0.2977 (37.70)\\
\hline
\end{tabular}
}
\end{center}
\vsm
\caption{\sl\small{Subspace Tracking results: Comparing altGDminMRI-ST1,  altGDminMRI-ST2, and onlineST algorithms for the Speech and the UnCardPerf sequences retrospectively undersampled using 16, 8, 4 radial lines and different choices of mini-batch size $\alpha$.
}}
\label{table_st_multicoil}
\end{table}

\begin{figure*}[t!]
\vspace{-0.25 cm}
\begin{center}
\includegraphics[width=10cm]{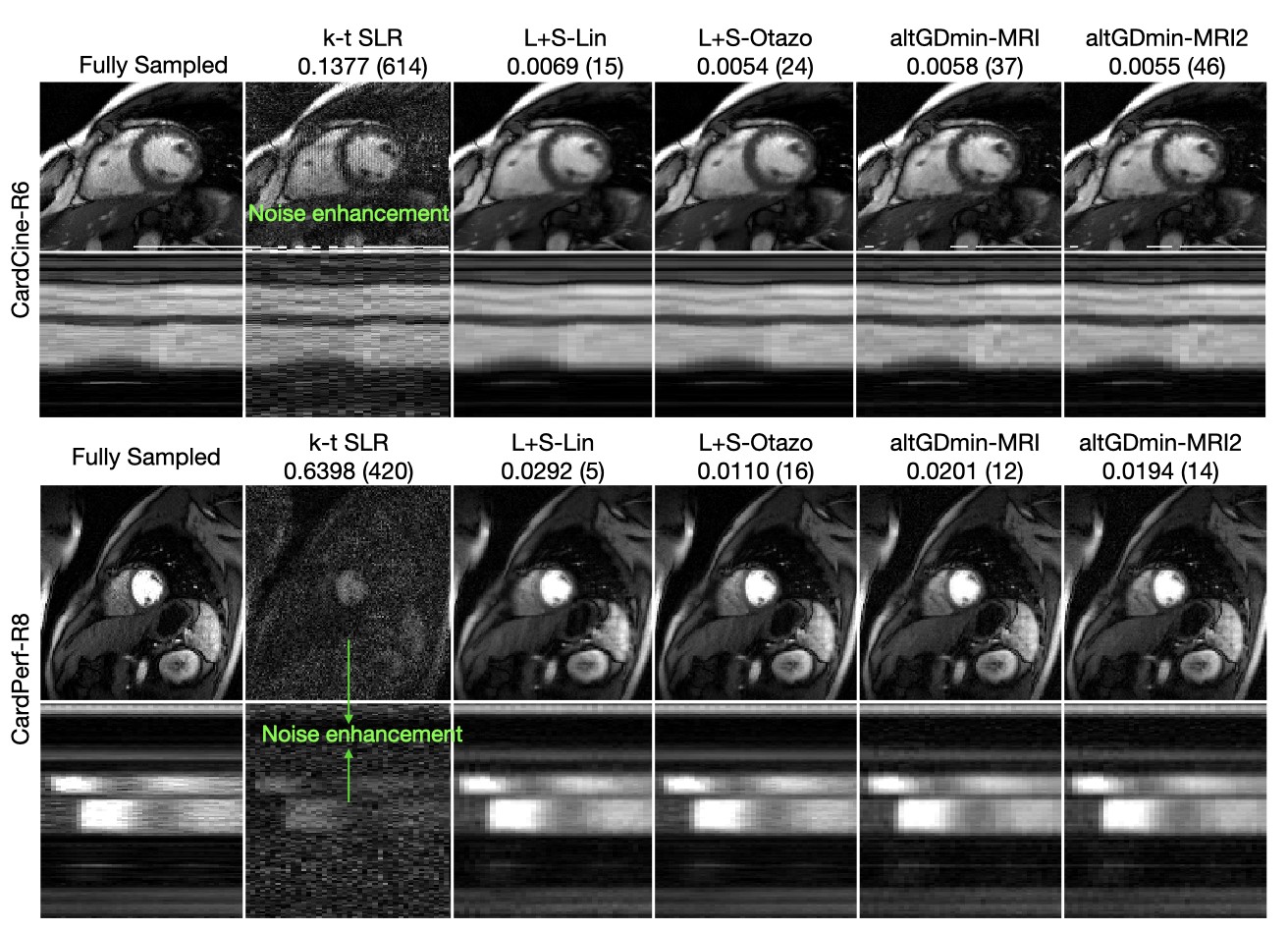}
\end{center}
\vsm
\caption{\sl\small{Retrospective, Cartesian, CardPerf and CardCine: Comparisons of reconstruction algorithms on CardPerf-R8 and CardCine-R6 datasets. In row 1 and row 3, we show one original frame (14th frame) and its reconstructions. In row 2 and row 4, we show the corresponding time profile images.  The chosen cut line is shown in Fig. \ref{allrecons_mc}.
{\bf Error (recon time)} of each algorithm is reported below the algorithm name. Observe that kt-SLR has considerable noise enhancement in both the datasets. AltGDmin-MRI2 (proposed) provides qualitatively similar results to L+S-Lin and L+S-Otazo.
}}
\label{cartesian_retro}
\vsm
\end{figure*}

\begin{figure*}[t!]

\begin{center}
\includegraphics[width=10cm]{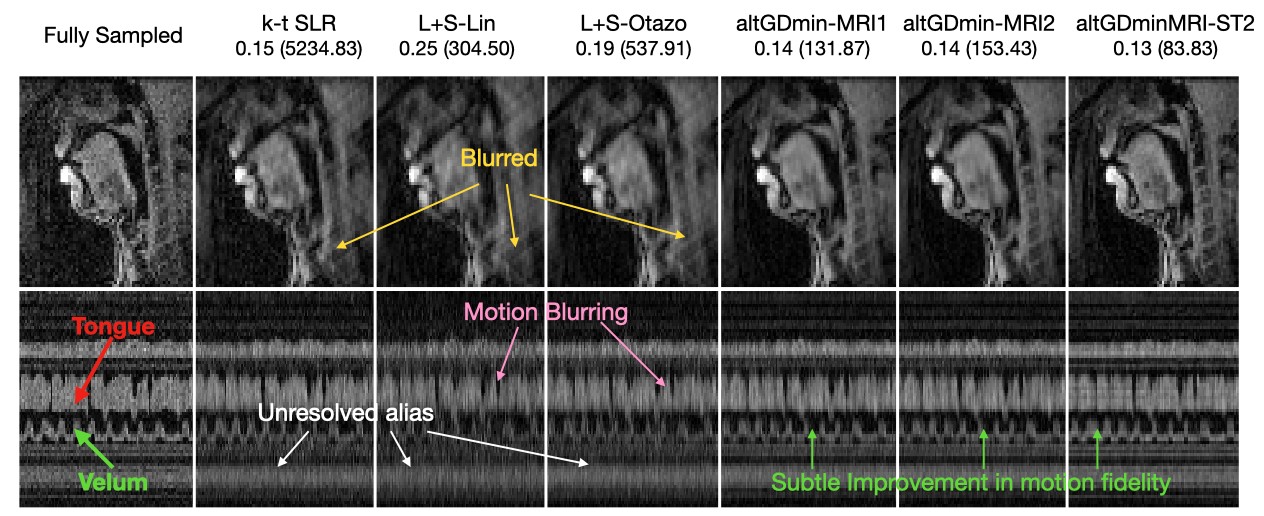}
\end{center}
\vsm
\caption{\sl\small{Retrospective, Pseudo-radial (4 radial lines), Speech:
In row 1, we show one original frame and its reconstructions. In row 2, we show the time profile image. This is a cut through the tongue and velum depicting the motion of these articulators. Only 100 out of 2048 image frames are shown for the sake of brevity.   The chosen cut line is shown in Fig. \ref{allrecons_mc}. {\bf Error (recon. time)} of each algorithm is reported below the algorithm name.
Observe that k-t-SLR, L+S-Lin, and L+S-Otazo reconstructions have motion blurring and/or alias artifacts. In contrast, altGDmin-MRI2  reconstructions even with smaller batch sizes produce superior reconstructions. Last column shows a subspace tracking result.
}}
\label{speech_4_mc_retro}
\vsm
\end{figure*}

\begin{figure*}[t!]
\begin{center}
\begin{subfigure}[t]{0.5\linewidth}
\hspace{-0.7 in}
\includegraphics[width=12 cm]{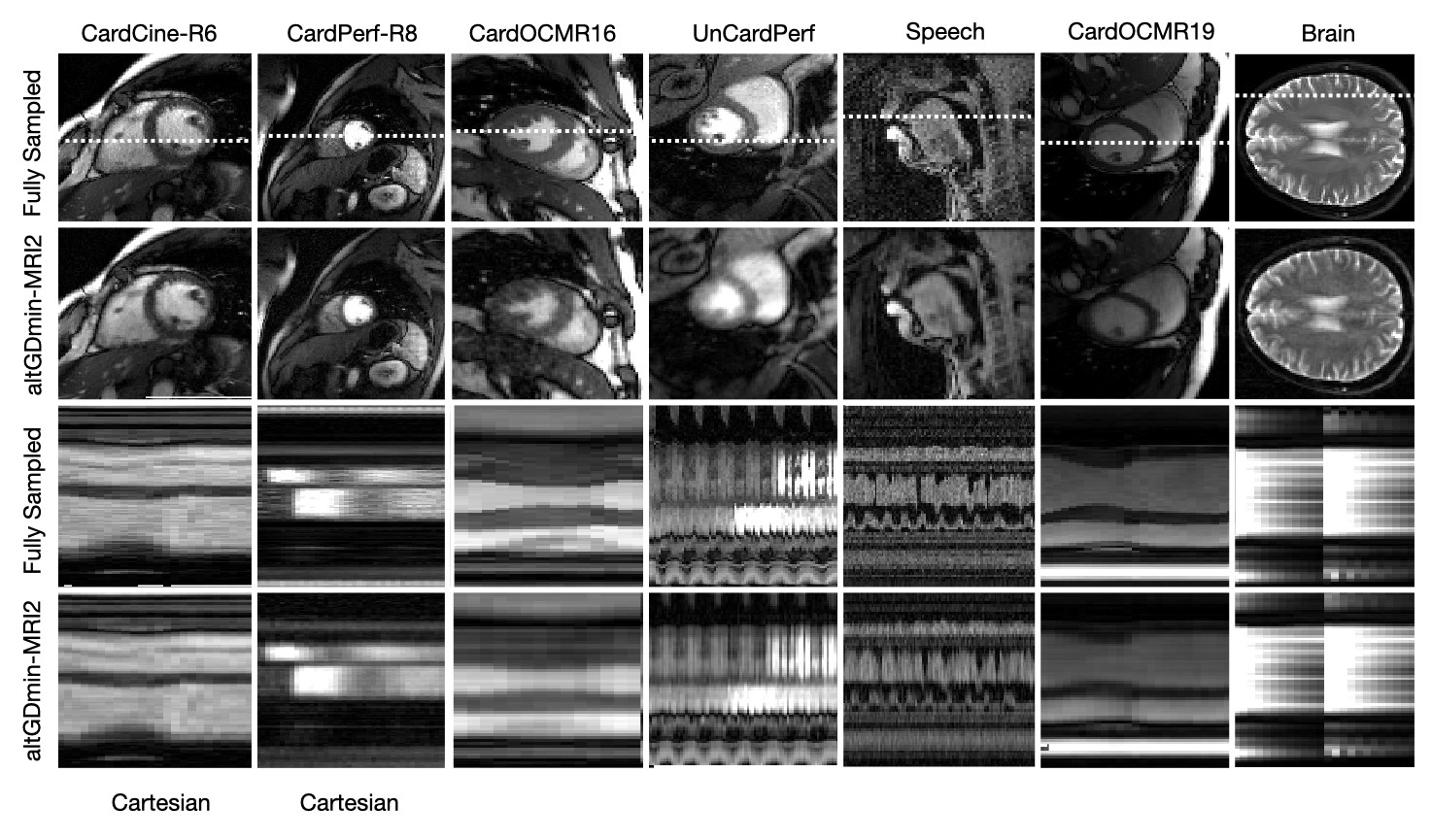}
\vspace{-0.35 in}
 \caption{\sl\small{All datasets with 4 radial lines} }
\end{subfigure}
\end{center}
\begin{center}
\begin{subfigure}[t]{0.5\linewidth}
\hspace{-0.7 in}
\includegraphics[width=12cm]{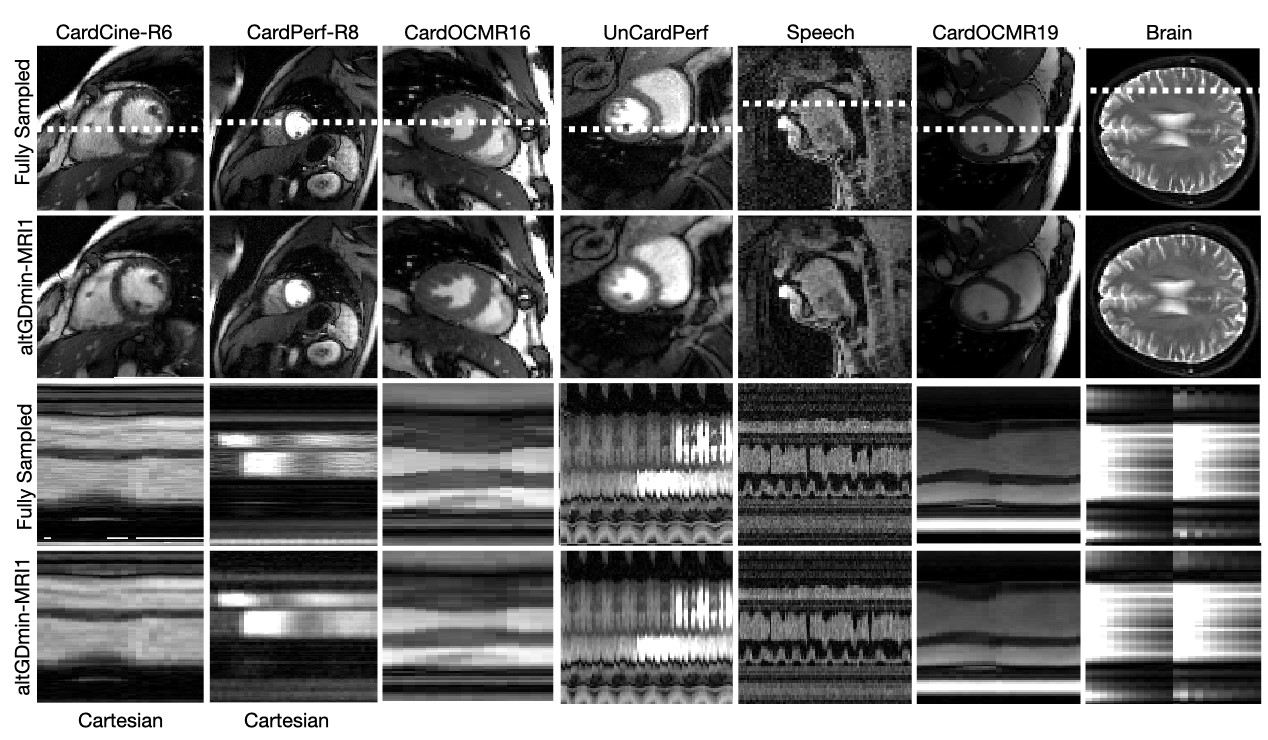}
\vspace{-0.35 in}
 \caption{\sl\small{All datasets with 16 radial lines} }
\end{subfigure}
\end{center}
\vsm
\caption{\sl\small{Retrospective, Pseudo-radial, All datasets: This figure is organized differently than the previous ones. Row 1 shows one original (fully sampled) image for all datasets, row 2 shows reconstruction using only AltGDmin-MRI2. Row 3 and row 4 are the original and reconstructed time profile images. Observe that it gives good results in all applications without any parameter tuning. Comparing Fig(a) and Fig(b, we observe that with 16 radial lines the blurring effect in the recons reduced.
}}
\label{allrecons_mc}
\vsm
\end{figure*}


\begin{figure*}[t!]
\begin{subfigure}[t]{0.5\linewidth}
\begin{center}
\includegraphics[width=8cm]{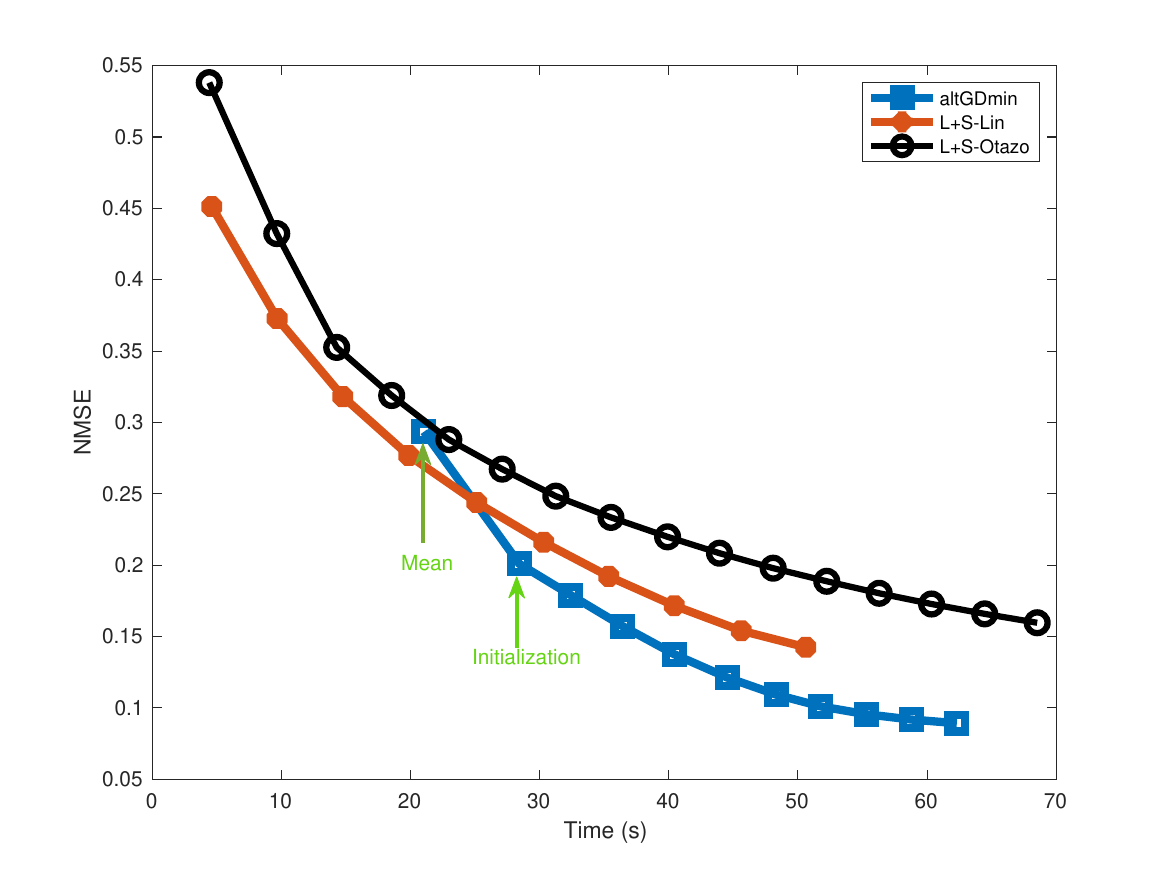}
 \caption{\sl\small{Error-Time: UnCardPerf } }
 \end{center}
\end{subfigure}
\begin{subfigure}[t]{0.5\linewidth}
\begin{center}
\includegraphics[width=8cm]{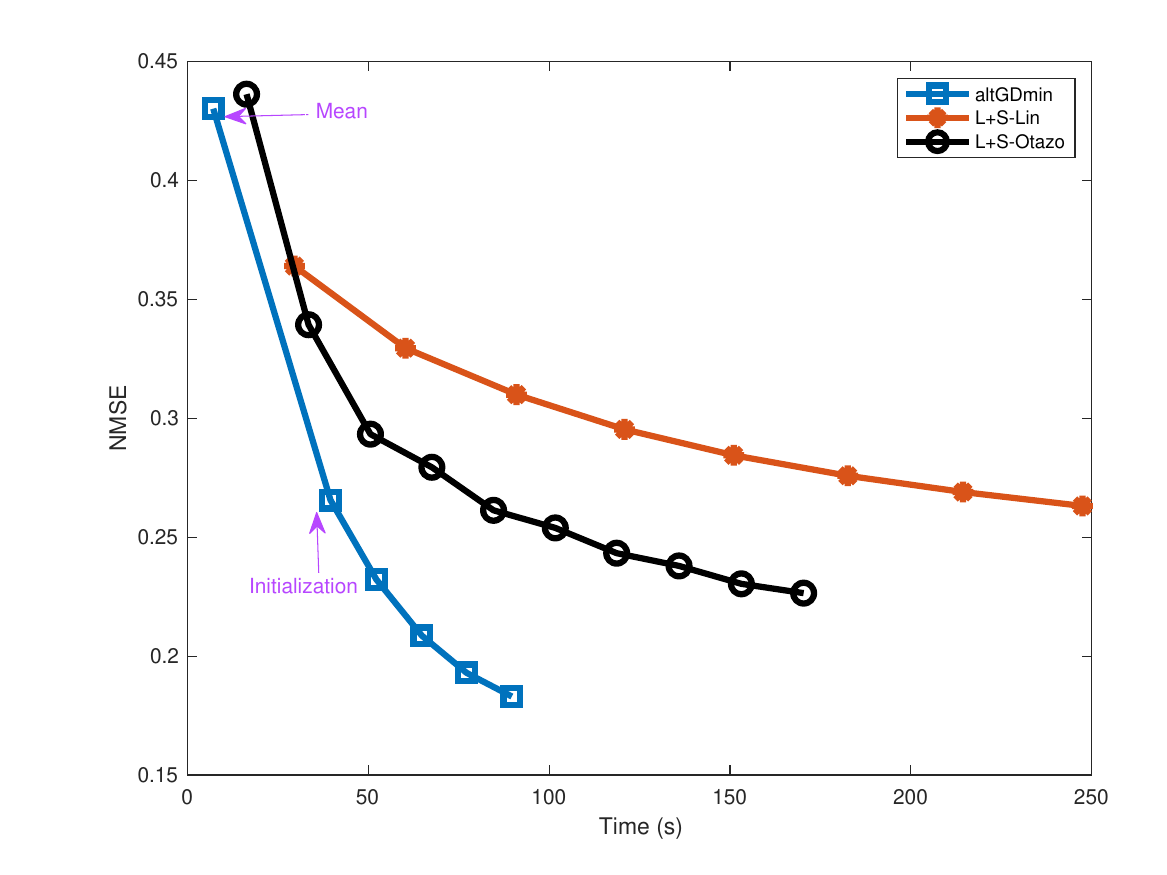}
\caption{\sl\small{Error-Time: Speech}}
\end{center}
\end{subfigure}
\caption{\sl\small{ 
We plot the  Error after each iteration t (y-axis) and the time taken until iteration t (x-axis) for the UnCardPerf and Speech datasets for alGDminMRI (without the MEC steps), L+S-Lin, and L+S-Otazo.
Observe that altGDmin-MRI converges fastest.
}}
\label{error_plot_cardiac_ungated}
\vsm
\end{figure*}

\begin{figure*}[t!]
\begin{center}
\includegraphics[width=9cm]{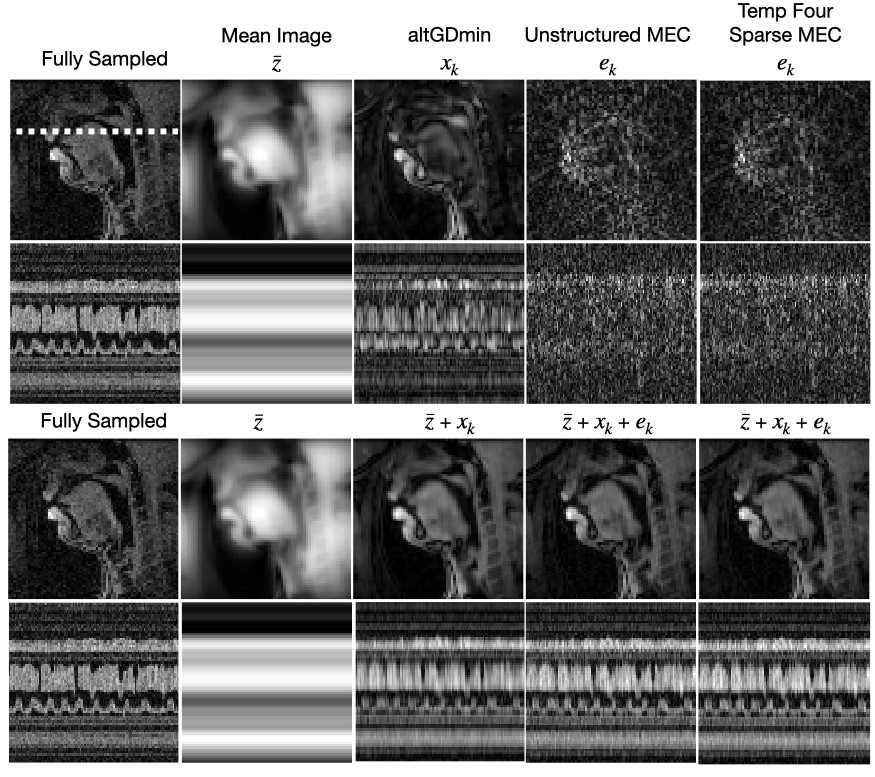}
\end{center}
\vsm
\caption{\sl\small{Demonstrating the utility of each step of our algorithms. In row 1, we show one original frame (1470th frame) and the estimates of each step: the mean estimate $\zbar$, the LR estimate $\x_k$ obtained using altGDmin, and the modeling error estimates $\e_k$ under our two models (unstructured and sparse in temporal Fourier domain). In row 2, we show the corresponding time profile images. In row 3, we show how each component improves image quality by showing $\zbar$, $\zbar + \x_k$, and $\zbar+\x_k + \e_k$. In row 4, we show the corresponding time profile images. Observe that altGDmin contributes to the details of each image. The last MEC steps improves finer details.
}}
\label{altgdmin_com}
\vsm
\end{figure*}

\begin{figure*}[t!]
\begin{center}
\begin{subfigure}[t]{0.5\linewidth}
\hspace{-0.9 in}
\includegraphics[width=14 cm]{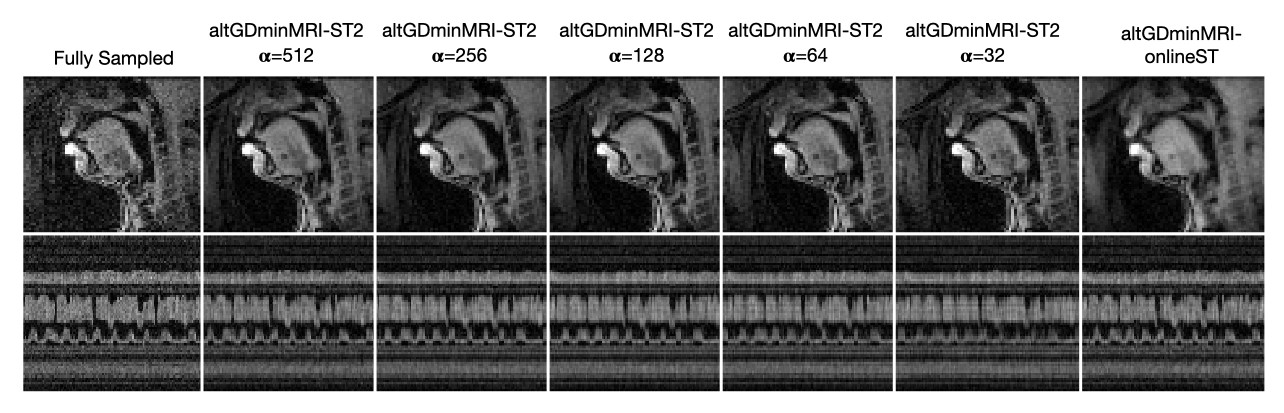}
\vspace{-0.2 in}
 \caption{\sl\small{Speech: 4 radial lines} }
\end{subfigure}
\end{center}
\begin{center}
\begin{subfigure}[t]{0.5\linewidth}
\hspace{0.2 in}\includegraphics[width=8 cm]{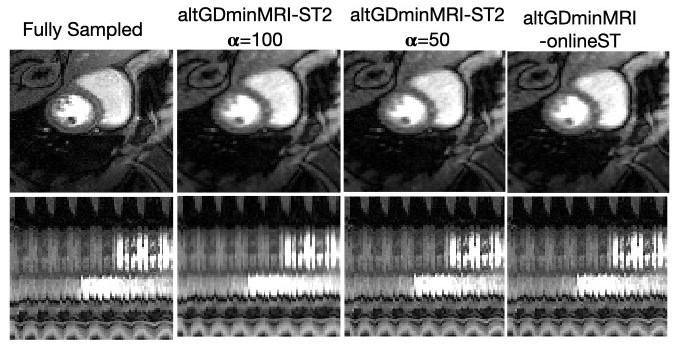}
\caption{\sl\small{Ungated Cardiac Perfusion (UnCardPerf): 16 radial lines}}
\end{subfigure}
\end{center}
\vsm
\caption{\sl\small{Subspace Tracking results:
Retrospective, Pseudo-radial , Speech(4 radial lines) and UnCardPerf (16 radial lines) reconstructed using altGDminMRI-ST2. In row 1, we show one original frame and its reconstructions. In row 2, we show the time profile image. As expected, smaller batch sizes produce faster reconstructions. With larger batch sizes,  the temporal sharpness (or motion fidelity) improve but only subtly.
}}
\vsm
\label{st}
\end{figure*}

\begin{figure}[t!]
\begin{center}
\vspace{-0.1 in}
\includegraphics[width=7cm]{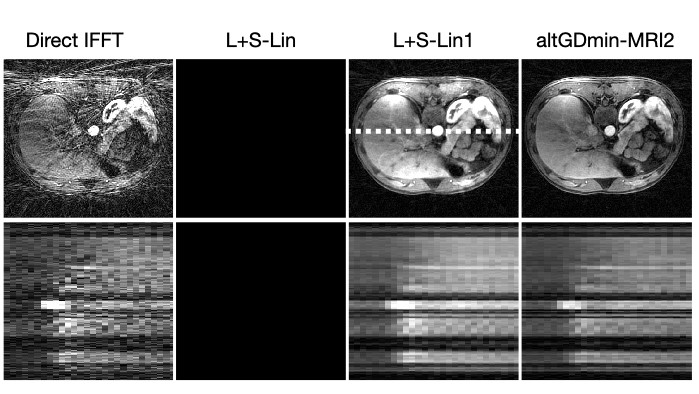}
\end{center}
\vsm \vsm 
\caption{\sl\small{Prospective, Radial, Abdomen: 
We compare our approach with direct iNUFFT (baseline) and with L+S-Lin. 
When running L+S-Lin with the cardiac perfusion parameters (the ones used in all earlier experiments), the algorithm completely fails, see column 2. Thus, we also implemented it using author-provided parameters for this dataset; we refer to this as L+S-Lin1 which is shown in column 3. This gives a good recovery similar to that of altGDmin-MRI2.
}}
\label{pros2}
\vsm
\end{figure}

\begin{figure*}[t!]
\centering
\begin{subfigure}{.5\textwidth}
  \centering
  \includegraphics[width=8 cm, height=4.4cm]{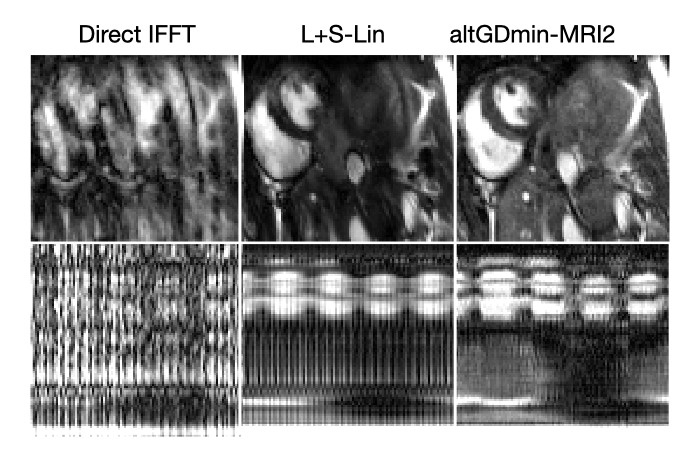}
\vspace{-0.3 cm}
\caption{OCMR dataset 1}
  \label{pros1}
\end{subfigure}%
\begin{subfigure}{.5\textwidth}
  \centering
  \includegraphics[width=7 cm, height=4cm]{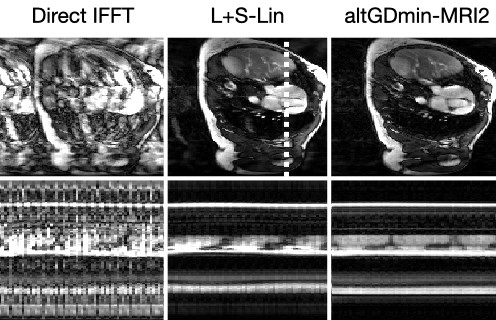}
\caption{OCMR dataset 2}
  \label{pros3}
\end{subfigure}
\vsm 
\caption{\sl\small{Prospective, Cartesian,  Cardiac (OCMR): In both figures, in row 1, we show one image for direct inverse FFT (IFFT) of undersampled kspace data (column 1), reconstructions using L+S-Lin (column 2) and our method, altGDmin-MRI2 (column 3). In row 2, we show the time profile images for the three reconstructions in the same order.
Observe that for both OCMR cardiac datasets, altGDmin-MRI2 reconstructions are qualitatively better compared to L+S-Lin.}}
\end{figure*}

%% file: revise2_dynmri_lrcs_expts.tex
\section{Experiments} \label{expts}
The code for all our experiments is posted at \url{https://github.com/Silpa1/comparison_of_algorithms}.
%
%
We show results on both retrospective and prospective datasets.
The retrospective undersampling is either 2D golden-angle based pseudo-radial\footnote{The angular increment between successive radial spokes is determined by the golden angle  (111.25 degrees) \cite{winkelmann2006optimal}. The starting point is changed over time so that the sampling masks are different for different images in the sequence. The golden angle ensures maximum incoherent kspace coverage over time.}  or 1D Cartesian undersampling using the sampling scheme of \cite{otazo2015low}. Radial sampling provides a way to non-uniformly undersample the 2D Fourier plane in such a way that more samples are acquired in the center of the 2D Fourier plan (low frequency regions in both dimensions). Directly using radially sampled data requires use of the computationally expensive non-uniform FT (NUFT) which makes the algorithm very slow. Previous research has shown negligible loss in image quality if the polar coordinates of radially undersampled data are regridded onto the Cartesian grid, followed by use of fast FT (FFT) algorithms in the reconstruction algorithms \cite{tian2017evaluation,benkert2018optimization,feng2018racer}.  
In fact, for CS-based methods, there is sometimes an improvement in reconstruction quality as well when using this type of regridded data (pseudo-radial data) and we observe this in some of our experiments too.
%
%
%

All experiments were conducted in MATLAB on the same PC. AltGDmin-MRI1 and AltGDmin-MRI2 were compared with (1) the three provable techniques --  mixed norm min (MixedNorm) \cite{lee2019neurips}, AltMin (changed for the current linear setting) \cite{lrpr_best} and basic AltGDmin \cite{lrpr_gdmin}; with 4 state-of-the-art methods from the LR-based MRI literature --  k-t-SLR (uses an L\&S model) \cite{lingala2011accelerated}, L+S-Otazo \cite{otazo2015low}, L+S-Lin \cite{lin_fessler}, and PSF-sparse \cite{zhao2012image,liang2007spatiotemporal}. 
For all comparisons, we used author provided code:
mixed norm min (MixedNorm): \url{https://www.dropbox.com/sh/lywtzc0y9awpvgz/AABbjuiuLWPy_8y7C3GQKo8pa?dl=0}, AltMin: \url{https://github.com/praneethmurthy/},
k-t-SLR: code was emailed by the author to us,  L+S-Lin: \url{https://github.com/JeffFessler/reproduce-l-s-dynamic-mri}, L+S-Otazo: \url{https://cai2r.net/resources/ls-reconstruction-matlab-code/},
PSF-sparse: \url{http://mri.beckman.illinois.edu/software.html}.

In all our experiments, one set of parameters was used. For our two methods, parameters were set as given earlier in the stepwise algorithms. For L+S-Lin, we evaluated it with using author-provided parameters for cardiac data and for PINCAT. Overall the cardiac parameters gave {\cblue reduced} errors and hence we used these in all experiments.  For L+S-Otazo and ktSLR also, author-provided cardiac parameters were used, since these gave the best results.
In the codes of kt-SLR, L+S-Otazo and L+S-Lin, the input k-space data is needed in a different format (the sequence of frequency locations is different, due to different uses of the ``fftshift'' in MATLAB). To deal with this, we converted all algorithms and ours so that all took input k-space data in the same format as L+S-Lin. 
L+S-Lin code has parameter settings that require the input k-space data to be normalized in a certain way. To generate the best performance for it, we did this normalization where needed. 


\subsection{Comparison with provably correct algorithms}\label{prove_res}
We compared altGDmin-basic and altGDmin-mean (altGDmin-MRI1 without the last MEC step) with MixedNorm \cite{lee2019neurips} and the AltMin algorithm of \cite{lrpr_icml,lrpr_best}, modified for the linear LRcCS problem (replace the PR step for updating $\b_k$'s by a simple LS step). 
Since these algorithms were designed and evaluated only for random Gaussian measurements, their code cannot be easily modified to handle large-sized image sequences (requires use of the fft operator to replace actual matrix-vector multiplications) or complicated MRI sampling patterns. Hence, for this experiment, we use a 30 x 30 piece of the PINCAT image sequence with 50 frames ($n=900,q=50$ and simulate (i) random Gaussian $\A_k$'s and (ii) random Fourier $\A_k$'s (the sampling mask is obtained by selecting $m$ 2D-DFT frequencies uniformly at random from all $n$ possible ones).
 We report the results in Table \ref{theory_algos1} for $m =n/10$. As can be seen, altGDmin and altGDmin-mean are more than 90-times faster than both altMin and MixedNorm. Also, altGDmin-mean has the lowest error. In the Gaussian setting, both altGDmin and altGDmin-mean have similar and small errors.  In the Fourier setting, since we are selecting random frequencies, if enough lower frequencies are not selected, the error is large. When the mean image is estimated and subtracted, the energy in the lower spatial frequencies is a lot lower. This is why use of mean subtraction (altGDmin-mean) significantly reduces the error in this case.

\subsection{Comparisons on retrospectively undersampled datasets} 
The error value that we report in this and later sections is normalized scale-invariant mean squared error  (N-S-MSE) computed as  follows $Error=  (\sum_{k=1}^q \dist^2 (\xstar_k,\hat\x_k) ) / \|\Xstar\|_F^2$ where $\dist^2 (\xstar,\hat\x) = \| \xstar - \hat\x \frac{\hat\x^\top \xstar}{\|\xhat\|^2}  \|^2$ is the scale invariant distance between two vectorized images with ``scale'' being a complex number. The reconstructed images can be complex-valued.
We also report the time taken to reconstruct the entire sequence. The reporting format is {\bf Error (Reconstruction Time in seconds)}. 

We used a total of 20 datasets: 2 datasets from  \cite{lin_fessler} which were retrospectively undersampled using Cartesian variable density random undersampling at reduction factors (R); R=8, R=6 respectively -- cardiac perfusion R8 (CardPerf-R8) and cardiac cine R6 (CardCine-R6); and 6 other applications that were retrospectively  pseudo-radially undersampled  with {\cred 4, 8 and 16 radial lines }
 -- brain-T2-T1-rho (Brain), free breathing ungated cardiac perfusion (UnCardPerf), a long but low-resolution speech sequence (Speech), two cardiac cine datasets from the OCMR database (CardOCMR16, CardOCMR19)\cite{chen2020ocmr}, and PINCAT. PINCAT data was single-coil, while all others were multi-coil.
Image sequence sizes: CardPerf  ($n=16384, q=40$), CardCine ($n=65536, q=24$), Brain ($n=16384, q=24$), Speech ($n=4624, q=2048$), UnCardPerf ($n=31104, q=200$), CardOCMR16 ($n=28800, q=15$), CardOCMR19 ($n=27648, q=25$), PINCAT ($n=16384, q=50$).  Dataset details are given in Appendix \ref{dataset_details}.

Error and time comparisons are reported in Table \ref{table_multicoil}. In its last row, we display Average Error (Average Reconstruction Time) with the average taken over the 20 previous rows. Visual comparisons are shown in Figs. \ref{cartesian_retro}, \ref{speech_4_mc_retro}, \ref{allrecons_mc}.
Observe that our approaches have the best errors and are also the fastest.
Both visually, and error-wise, AltGDmin-MRI2 has either the best or a close second-best reconstruction quality in all cases, while also being very fast. It is not always the fastest, but it is the fastest for long sequences and fast-enough for all. AltGDmin-MRI1 also has low errors (only slightly higher than MRI2), and is faster than MRI2. On the other hand, no other approach is consistently good across all 20 datasets.
Lastly, for the most undersampled (4 radial lines) case, our methods have much lower errors than all the others.
We also compare with the PSF-sparse algorithm of  \cite{zhao2012image} for one dataset, UnCardPerf. This is an improved version of the original kt-PCA method of \cite{liang2007spatiotemporal}. This approach does not work for any given sampling scheme. Hence, for it, we used the author provided ``k-t sampling'' code (Cartesian undersampling with certain k-space locations sampled at each time, and the rest of the locations being highly undersampled).
We changed its sampling rate parameter to make its undersampling factor, $\sum_k m_k/(nq)$,  similar to ours. 
From Table \ref{table_multicoil}, clearly, the error of PSF-sparse is much higher (38- and 12- times higher) in the 4 and 8 radial lines cases.%

In this table, we have also added results for one mini-batch size for altGDminMRI-ST2. As can be seen, in the speech sequence case, the recovery error is in fact smaller.
In the cardiac case,  there is marginal increase in recovery error. In both cases, the time taken is lesser than that of altGDmin-MRI2. More detailed evaluation is described below in Sec \ref{expts_st}.

\subsection{Error versus iteration time plot}
In Fig. \ref{error_plot_cardiac_ungated}, we plot the error after each iteration $t$ (y-axis) and the time taken until iteration $t$ (x-axis) for altGDmin-MRI without the MEC step (just altGDmin+mean), L+S-Lin and L+S-Otazo for the UnCardPerf and Speech datasets with 4 radial lines. Such a plot is more informative than error versus iteration since it allows one to both see the error decay with iterations and to also see the time taken for each iteration by each method.
The first marker in the altGDmin-MRI plot is time taken after the mean computation, the second marker also adds the time taken after the initialization step, the third also adds the time taken by first altGDmin iteration, and so on. All three algorithms have their own exit loop and maximum number of iterations and hence each plot ends at a different time.
For both the datasets, observe that altGDmin converges much faster than the other two compared methods. kt-SLR is not compared because it is much slower. 
From both figures, we can also observe that use of our initialization step is very useful, it helps reduce the error significantly. With just the mean and initialization steps, the normalized error gets reduced to 0.2 and 0.27 respectively.


\subsection{Effect of each step of altGDmin-MRI}
In Fig.\ref{altgdmin_com}, for one dataset, we show the output of each step of our two algorithms. As can be seen each of the 3 steps improves the reconstruction quality.


\subsection{Subspace tracking algorithms evaluation}\label{expts_st}
We evaluate these algorithms on the Speech and the UnCardPerf sequence pseudo-radially undersampled using 16, 8, and 4 radial lines. We use these two sequences since these have a larger value of $q$. This is needed to ensure that the mini-batch sizes are not too small or at least the first mini-batch size is large enough.
We provide the results in Table \ref{table_st_multicoil}.  We evaluate mini-batch ST (Algorithm \ref{gdmin_ST}), with $T_{max,1}=70$  and $T_{max,j}=5$ for $j>1$ for decreasing values of $\alpha$.  
Observe that for mini-batch sizes up to $\alpha \ge 64$, there is no appreciable increase in error. 
But the improvement in reconstruction time is very significant. It is much faster than any of the other algorithms compared in Table \ref{table_multicoil}.
%
%
We also compare with full online ST (Algorithm \ref{gdmin_onlineST}). 
In this case, there is a significant increase in error as the value of initial mini-batch $\alpha$ decreases. But the speed is even better.
We show visuals for different values of $\alpha$ in Fig. \ref{st}. As can be seen, the quality is as good as that of the full batch one. The reason is this approach is modeling a slowly changing subspace rather than a fixed one, and this can be a better assumption for speech sequences.

%
%

\subsection{Experiments on 3 prospectively under-sampled radial data} \label{prospec_expt}


 We compare altGDmin-MRI2 with L+S-Lin (the overall best algorithm in terms of performance and speed amongst all compared methods) and with the baseline reconstruction obtained using direct inverse Fourier Transform (FT), this uses zeros where data is not observed. Results are show in Fig. \ref{pros2}, \ref{pros1}, \ref{pros3}.
%
Our first dataset is a radially undersampled dynamic contrast enhanced (DCE) abdomen taken from \cite{otazo2015low,lin_fessler}. {\cred
The k-space data dimensions in the dataset were: 384 read out points, 21 radial spokes (with golden angle based angular increments) per frame, 28 time frames, and 7 virtual coils after PCA based coil compression. Here 21 spokes per frame means the total $21*28=588$ spokes were arranged into 28 time frames with the first 21 spokes forming the first time frame, the next 21 forming the next time frame and so on.
}
For this dataset, we modified our code to use the non-uniform Fast FT (NUFFT) code from \cite{jeff_nufft} to replace FFT. 
From Fig. \ref{pros2}, notice that, in the altGDmin-MRI2 (atGDmin2) reconstruction, we can observe the contrast uptake dynamics through the liver. The other blood vessels are well resolved too. L+S-Lin (with the parameters used in all previous experiments) failed completely. It returns a black image. So we used the author provided parameters to see if that works, we label it L+S-Lin1. This provides qualitatively similar results to ours.
%

Our next dataset is a Cartesian undersampled breath held cardiac cine from the OCMR database with k-space dataset dimensions 384 read out points, 14 phase encode lines per time frame, 137 time frames and 18 coils. From Fig. \ref{pros1}, observe that L+S-Lin reconstruction has motion blurring and/or alias artifacts. In contrast, altGDmin2 result is good comparatively.%
%
%
Our last dataset is another Cartesian undersampled cardiac dataset from the OCMR database with k-space data dimensions 384 read out points, 16  phase encode lines per time frame, 65 time frames and 34 coils. In Fig. \ref{pros3}, we compare the reconstructions.  Both L+S-Lin and altGDmin-MRI2 reconstructions are good. 

%% file: revise2_dynmri_lrcs_discussion.tex
\section{Discussion}\label{discuss}
We first explain why our algorithms are ``general'', memory-efficient, and fast. Next, we provide a summary of our experimental conclusions, a discussion of the most related works, and of our subspace tracking methods (ST). We end with describing the limitations of our work and ways to improve it.

\subsubsection{AltGDmin-MRI methods are ``general''}
This is because these have only a few parameters and are not very sensitive to their choices. Moreover, our goal  was to develop a {\em single algorithm, with one fixed set of parameters}, that provides a good enough performance across a wide range of applications, sampling schemes, and sampling rates, while also being very fast; and not necessarily the best one for each case. Hence we do not do any application-specific tuning. 

The reason our methods have only a few parameters is two-fold. First, the LR model does not require any parameter except the assumed rank (or parameters for the algorithm to estimate the rank). This is unlike sparsity or structured sparsity models, which require either picking the most appropriate sparsifying basis or dictionary or learning one, and in either case there are many parameters that need to be carefully set to pick the best basis/dictionary. 
Second,
AltGDmin is a simple GD based algorithm. Besides rank, its only other parameters are the GD step size $\eta$ and the maximum number of iterations $T_{max}$, along with a loop exit threshold $\epsilon_{exit}$. The algorithm is not sensitive to the choice of $T_{max}$ as long as it is large enough.
AltGDmin-MRI1 is its modification that can be understood as assuming a 3-level hierarchical LR model that (i) first estimates the baseline/mean image across all frames (approximately the  $r=1$ case), and computes the measurement residual by removing this estimate; (ii) next it uses the residual as input to auto-altGDmin ($r=\rhat$ case, where $\rhat$ is the automatically estimated rank), and (iii) finally it estimates the residual error in the above mean + LR model column-wise (this is the $r=\min(n,q)$ case). In case of altGDmin-MRI2, this last residual error is assumed to be temporally Fourier sparse.
The mean computation step (LS problem solved using the Stanford CGLS code) and the MEC steps also require only two parameters each, while being sensitive to only one of them: loop exit tolerance and maximum number of iterations. The ISTA algorithm used in case of altGDmin-MRI2 needs one more parameter - the threshold for soft thresholding.


\subsubsection{Memory-efficiency}
The effect of memory complexity is not very evident in this paper since we only do single-slice imaging, but will be when working with dynamic multi-slice imaging. In that case, $n$ would be the number of voxels in each volume (all slices at one time) with $q$ still being the sequence length.
AltGDmin uses the $\X = \U \B$ factorization, with $\U$ and $\B$ being matrices with $r$ columns and rows respectively. Here $r$ is the assumed (low) rank.  Storing and processing $\U,\B$ requires memory of size only $\max(n,q) r$ instead of $nq$. Our approach for estimating $r$ caps its value at $r \le \min(n,q,m)/10$.
The initial mean computation step estimates an $n$-length vector $\zbar$ using GD (CGLS code); it can be made memory-efficient by using a for-loop to compute the gradient sum at each iteration. For AltGDmin-MRI1, the last MEC step is done individually for each $\estar_k$. Thus both these steps have memory complexity of order $n$ only.
The MEC step of AltGDmin-MRI2 can be made memory-efficient by processing each row separately using a for-loop over all $n$ rows. This step thus has memory complexity of order $q$.
Thus the overall memory complexity of both  AltGDmin-MRI1 and  AltGDmin-MRI2 is order $\max(n,q) r$ with $r \ll n,q$, while that of {\cred most other LR-based methods (except AltMin and PSF-sparse)} is $nq$.



\subsubsection{Time complexity and speed}
The most computationally expensive part of both algorithms is altGDmin. The two expensive steps of this algorithm are (i) computing the gradient w.r.t. $\U$, and (ii) computing $\A_k \U$ for the LS step to estimate $\b_k$s. When implemented for the MRI setting using the 2D-FFT operators, $\A_k \U$ requires computing $r$ 2D-FFTs for $n_1 \times n_2$ images with $n=n_1 \cdot n_2$. 
One 2D-FFT needs time of order $n_2 n_1 \log n_1 + n_1 n_2 \log n_2 \le 2 n \log n$. Thus this step needs time of order $nr \log n$. The gradient computation needs $q$ 2D inverse FFTs, thus its cost is $nq \log n$. Since $r \le q$, the overall cost is order $n q \log n$ per iteration. Without provable guarantees for the MRI setting, we cannot say anything theoretically about the number of iterations required. 
From our experiments (see Fig.  \ref{error_plot_cardiac_ungated} and Tables  \ref{theory_algos1}, \ref{table_multicoil}), our algorithm error decays faster than that of all the other compared approaches.


\subsubsection{Summary of experiments}
In the highly undersampled setting of only 4 radial lines (32 times acceleration),  AltGDmin-MRI2 and AltGDmin-MRI1 have much lower errors and better visual recon quality than all compared methods, while also being the fastest. In all cases, on average, the AltGDmin-MRI2 errors and time taken are still the best (lowest).
Our Subspace Tracking based mini-batch modifications are even faster, while providing almost comparable, or better, quality reconstructions for batch sizes $\alpha \ge 64$. For this reason, these can only be used on longer sequences.
Notice that the last model error correct (MEC) step of altGDmin-MRI1 uses a maximum of only 3 iterations, while that of altGDmin-MRI2 uses a maximum of 10 iterations. As noted by an anonymous reviewer, this may be one reason for the latter having slightly better performance. We tried using 10 maximum iterations also for MEC of altGDmin-MRI1 (not shown); with this, its computed error does reduce further to almost the same level as MRI2. However, for a some cases, the visual quality of MRI2 still is better.

\subsubsection{Discussion of compared methods and DL-based methods}
MixedNorm, AltMin, and kt-SLR are much slower compared with AltGDmin-MRI.
For many of the pseudo-radial datasets, kt-SLR has the lowest errors and visual performance, with that of AltGDmin-MRI2 being either as good or only slightly worse.
But kt-SLR has very large errors for the Cartesian undersampled ones and consequently its average error is large.  A possible reason for this is that its parameters have been tuned for the pseudo-radial sampling.
On the other hand, L+S-Otazo and L+S-Lin have low errors for the two Cartesian undersampled datasets (both of these were used in their papers); but have larger errors for most of the pseudo-radial 4 and 8 radial lines (highly undersampled)  datasets. The  likely reason again is similar. 

We also compared with the PSF-sparse algorithm of \cite{Liang2012} for one dataset; see Table \ref{table_multicoil}. This only works with the kt-sampling scheme developed by the authors, so we used this sampling for it while changing the code parameters to ensure comparable $ mc \sum_k m_k/(nq)$ (comparable acceleration factor).  PSF-sparse first estimates the row space of the unknown image sequence matrix using the data at the frequency locations fully sampled along time. This means it estimates the row span of $\B$ first. Next, it estimates the column space, span of $\U$, by minimizing the error w.r.t. all observed data while also imposing a (temporal Fourier) sparsity constraint. The first step needs sufficient number of low frequency samples for accurate recovery. This is why, when $\sum_k m_k$ is small, either the first step recon is bad or there are almost no samples left to get a good estimate at all frequencies. In contrast, our algorithms simultaneously estimate $\U, \B$ using multiple alternating iterations initialized using a carefully designed spectral initialization for $\U$.  

Supervised DL methods need a lot of training data. Hence for most MRI applications (except breath held cardiac or ECG gating), these are designed for image-based reconstruction and cannot model the spatiotemporal correlations across the sequence. Consequently, their performance on dynamic MRI is often worse than that of LR or LR+S based methods which do not need any training data.
Moreover, the energy cost for training the DLs for entire image sequences can be prohibitive. 

On the other hand, the new unsupervised DL methods can model the spatiotemporal correlations without training data, but these are slower on query data by orders of magnitude. The reason is these do not use a pre-trained network but instead train the deep network on the query data. As an example, to reconstruct a typical speech sequence with $n = 100^2 = 10000$ pixels and $q = 500$ frames, this class of approaches needs 45mins to an hour on a GPU. Our algorithm only needs 1-2 minutes for a similar dataset. Also, these need careful hyperparameter tuning (cannot be used for different MRI applications without tuning) while our methods do not.



{\cred
\subsubsection{Clarification that no binning is needed in this work} 
None of our proposed approaches (not even the subspace tracking ones) need raw data sorted into various predefined temporal phases (such as defining cardiac phases) or ``bins''. 
As an example, for the cardiac cine sequence, we did not need knowledge of which frames are pre-contrast and which are post-contrast. 
%
}

\subsubsection{Practical utility of Mini-batch and Online Subspace Tracking (ST)}
Besides operating in mini-batch mode, the algorithm speed of mini-batch ST is also much faster, and, from our experiments, the increase in recovery error is insignificant for mini-batch size $\alpha \ge 64$. In the speech sequence, the error actually decreases when using ST. The reason is this is a very long sequence and the subspace likely changes over time. The ST method tracks this change. With online ST, the reconstruction quality does suffer. But its speed is very fast, and the algorithm works in true real time mode after the first mini-batch of $\alpha_1$ frames is processed. Hence, in practice, a combination of the two approaches would be the most useful: in online mode, obtain real-time reconstructions which are very fast but often not very accurate; and follow it up with mini-batch updates (after the mini-batch has arrived) that are much more accurate.

{\cred 
Because we do not need any binning, and because the ST approaches provide a reconstruction as soon as a small mini-batch of time frames are acquired, these would be suitable for a variety of real-time ungated type of applications, where the raw data is continuously being acquired without an a-priori definition of temporal phases. Example  applications include  free breathing ungated cardiac cine, real-time dynamic MRI of vocal tract shaping during speech production, free breathing dynamic contrast enhanced MRI (also see example in Fig. 8). There is a need to have a fast on-the-fly reconstructions to inspect quality of the dynamic reconstructions. For example, in dynamic speech MRI, low latency reconstructions are useful to adjust for localization planes,  adjust center frequency to minimize off-resonance artifacts,  and to visualize articulatory movements in biofeedback type experiments. Similarly, in real-time ungated free breathing cardiac MRI experiments, a fast reconstruction without a prior definition of cardiac phases allows one to visualize arrhythmic events on the fly.
}

\subsubsection{Limitations and how to tune parameters to tailor to applications}
As can be seen from our results, the reconstruction performance is not the best for all applications. In a few cases, there is some visible blurring. The reason is our goal was to show what our algorithm can achieve using a single set of parameters. 
The blurring can be reduced by increasing the rank $\rhat$ that is used in AltGDmin while still ensuring it is sufficiently smaller than $\min(mc \min_k m_k, n, q)$; and/or increasing the number of MEC step iterations. Making the loop exit criterion more robust can also help, e.g., instead of exiting after only two consecutive estimates of $\U$ are close in subspace distance, one could exit if this happens consecutively for a few iterations. The algorithm speed can be improved further  by using a variable step size $\eta_t$ for the GD step: use larger values in the initial iterations and reduce it over time.
Lastly, the initial mean computation step of altGDmin-MRI1 and altGDmin-MRI2 can be made more robust by using truncation similar to that used in the initialization of altGDmin.%

For both mini-batch and online ST methods, we need to initialize using a mini-batch. From our experiments, the first mini-batch needs to be at least 32-64 frames long. It is possible to replace even the first mini-batch step by a fully-online one if we replace the GD for updating $\U$ by stochastic GD that only uses the gradient w.r.t. the data term for the current $\y_k$. However, the tradeoff will be a worsened reconstruction quality. This approach will be explored in future work.
Moreover, our ST methods are not very robust to outliers, e.g., due to a deep breath by the subject. One way to make them somewhat robust is to re-initialize every so often.  The second solution is to develop an L+S model based algorithm. 



\section{Conclusions and future work}\label{conclude}

We developed a set of fast, memory-efficient, and ``general" algorithms for accelerated/undersampled approximately LR dynamic MRI. 
Here, ``general" means that it works with the same set of parameters for multiple MRI applications, sampling schemes and rates. We also developed an altGDmin-MRI-based subspace tracking solution that operates in mini-batch mode and provides comparable reconstruction quality while being even faster.
Unlike the supervised DL methods, our algorithms do not need any training data, and also do not need the hours or days of training time and computational power. Unlike the newer unsupervised DL based methods, our methods are orders of magnitude faster. In experimental comparisons with the best known existing LR, L+S or L\&S based methods (kt-SLR, L+S-Otazo, L+S-Lin, PSF-sparse, AltMin, MixedNorm), on average, our methods have the best reconstruction quality, and are also the fastest. 
%
Future work will explore reconstruction of 3D+t data (multi-slice dynamic imaging) using the proposed approach and also try to develop tensor-based modification of our ideas. A second goal will be to design a fast and memory-efficient altGDmin-based algorithm for the L+S model. We will also try to design a truly online ST method that does not need mini-batch initialization.%

%% file: revise_dynmri_lrcs_appendix.tex
\section{Dataset details - Retrospective case}\label{dataset_details}  

Fully sampled multi coil datasets acquired on a Cartesian grid in different dynamic MRI applications were retrospectively downsampled using a 1D Cartesian undersampling mask, or a 2D pseudo radial sampling mask. Details of the full-sampled datasets are provided below. Using standard format from MRI literature, the dimensions are stated as $k_x \times k_y \times q \times mc$ where $k_x \times k_y$ is the image size, thus $n = k_x \cdot k_y$; $q$ is the number of frames; and $mc$ is the number of coils used.
For our experiments, the raw data of datasets in (3)-(6) below were coil-compressed into 8 virtual coils. The virtual coil sensitivity maps were estimated using the Walsh eigen decomposition algorithm \cite{walsh2000adaptive} from time averaged undersampled data. For datasets in (1), and (2), we did not perform any coil compression, and used the author provided coil sensitivity maps. For these, we used the author provided retrospectively undersampled data which was undersampled using 1D variable density random undersampling mask at reduction factors (R); R=8, R=6 respectively.

\begin{enumerate}
\item A breath held cardiac perfusion dataset with dimensions  $k_x \times k_y \times q \times mc =128\times 128 \times 40 \times 12$ previously used in \cite{otazo2015low,lin_fessler}.

\item A breath held cardiac cine dataset with dimensions of $k_x \times k_y \times q \times mc =256\times 256 \times 24 \times 12$ previously used in \cite{otazo2015low,lin_fessler}.

\item A low spatial but high temporal resolution speech data set with dimensions of $k_x \times k_y \times q \times mc =68\times 68 \times 2048 \times 16$. This data was  acquired at the University of Iowa on a healthy volunteer producing a variety of speech sounds
including uttering interleaved vowel and consonant sounds
(za-na-za-loo-lee-laa), and counting numbers at the subjects
natural speaking rate. Since the subject was speaking at the natural rate, the spatial
resolution was compromised to $3.4 \times 3.4 mm^2$ and the temporal
resolution was maintained at 70 ms/frame to capture the
motion of the articulators.

\item A brain multi-parameter (T2, and T1-rho) mapping
dataset with dimensions of   $k_x \times k_y \times q \times mc =128\times 128 \times 24 \times 12$ previously used in \cite{otazo2015low,lin_fessler}. This was acquired on a normal
volunteer at the University of Iowa. Spin lock times and echo
times were varied to estimate T2, and T1-rho (T1 in rotating
frame) time constants from the resulting multi-contrast (or
loosely dynamic) image frames.

\item One free breathing ungated cardiac perfusion dataset, previously used in \cite{lingala2014deformation} with dimensions  $k_x \times k_y \times q \times mc =288\times 108 \times 200 \times 32 $. This dataset
had both perfusion dynamics and motion dynamics due to
cardiac and breathing motion. The dataset was acquired at the University of Utah with TR/TE=2.5/1ms; saturation recovery
time=100 ms.

\item Two representative breath held short axis cardiac cine datasets from the OCMR database \cite{chen2020ocmr} with dimensions respectively $k_x \times k_y \times q \times mc =192\times 144 \times 25 \times 38 $; and $k_x \times k_y \times q \times mc =192\times 150 \times 15 \times 30 $.

\item PINCAT: Free breathing PINCAT perfusion
phantom containing dynamics due to perfusion uptake in
the heart as well as heavy breathing motion. This phantom
was previously used in the works of \cite{lingala2011accelerated,lin_fessler}.
\end{enumerate}